%% file: Long-Version-V6.tex
\documentclass[11pt,a4paper]{article}
\usepackage{amsmath,latexsym,fullpage,amssymb,color}
\usepackage{epic,eepic,ifthen,graphics,epsfig}
\usepackage[english]{babel}
\usepackage{times}
\usepackage{float}
\usepackage{xspace}
\usepackage[T1]{fontenc}
\newlength {\squarewidth}

\newcommand{\toto}{xxx}

\newenvironment{lemma-repeat}[1]{\begin{trivlist}
\item[\hspace{\labelsep}{\bf\noindent Lemma~\ref{#1} }]}%
{\end{trivlist}}

\newenvironment{theorem-repeat}[1]{\begin{trivlist}
\item[\hspace{\labelsep}{\bf\noindent Theorem~\ref{#1} }]}%
{\end{trivlist}}

\newcounter{linecounter}
\newcommand{\linenumbering}{\ifthenelse{\value{linecounter}<10}
{(\arabic{linecounter})}{(\arabic{linecounter})}}

\newcommand{\resetline}[1]{\setcounter{linecounter}{0}#1}
\renewcommand{\thelinecounter}{\ifnum \value{linecounter} > 
9 \else \fi\arabic{linecounter}}
\newfloat{algorithm}{thp}{lop}
\floatname{algorithm}{Algorithm}
\newfloat{construction}{thp}{lop}
\floatname{construction}{Construction}

\newfloat{sidebar}{thp}{lop}
\floatname{sidebar}{Sidebar}

\newcommand{\Xomit}[1]{}
\newcommand{\tto}{{\sf to}}  
\newcommand{\ssend}{{\sf send}}  
\newcommand{\bbroadcast}{{\sf broadcast}}  
\newcommand{\wwait}{{\sf wait}}  
\newcommand{\wwrite}{{\sf write}}  
\newcommand{\rread}{{\sf read}}  
\newcommand{\return}{{\sf return}}

\newcommand{\tup}[1]{\langle #1 \rangle}
\newcommand{\STATE}{{\mathit{STATE}}}  
\newcommand{\BOARD}{{\mathit{BOARD}}}  
\newcommand{\REG}{{\mathit{REG}}}  

\begin{document}

\title{\bf Mastering Concurrent Computing \\Through Sequential Thinking:\\
  A Half-century Evolution}

\author{Sergio Rajsbaum$^{\dag}$,
        Michel Raynal$^{\star,\circ}$\\~\\
$^{\dag}$Instituto de Matem\'aticas, UNAM, Mexico\\
$^{\star}$Univ  Rennes IRISA, 35042 Rennes, France \\
$^{\circ}$Department of Computing, Hong Kong Polytechnic University \\        
{\small {\tt rajsbaum@im.unam.mx~~~~~ raynal@irisa.fr}}
}

\date{}

\maketitle


\begin{abstract}
Concurrency, the art of doing many things at the same time is slowly
becoming a science. It is very difficult to master, yet it arises all
over modern computing systems, both when the communication medium is
shared memory and when it is by message passing.  Concurrent
programming is hard because it requires to  cope with many possible,
unpredictable behaviors of communicating processes interacting with
each other.  Right from the start in the 1960s, the main way of
dealing with concurrency has been by reduction to sequential
reasoning. We trace this history,
and illustrate it through several examples, from early ideas based on
mutual exclusion, passing through consensus and concurrent objects,
until today ledgers and blockchains. We conclude with a discussion on
the limits that this approach encounters, related to fault-tolerance,
performance, and inherently concurrent problems.
%

.~ 
\end{abstract}

\newpage

\begin{flushright}
{\it
I must appeal to the patience of the wondering reader, \\suffering as I am 
 from the sequential nature of human communication.}\\
E. W. Dijkstra, 1968~\cite{D68}
\end{flushright}


\section*{INTRODUCTION}

\paragraph{Sequential reasoning is natural and easier.}
The human brain behaves as a  multiprocessor computer,
which performs many tasks simultaneously, naturally and frequently.
However, despite the fact  we are good at processing parallel information, 
it is difficult to be aware of the activities we perform concurrently, and
when we try to raise awareness, we end up distracting ourselves and
reducing the quality of what we are doing. Only after intense training
can we, like a musician,  conduct   several
activities simultaneously.

 It is infinitely easier to reason sequentially, doing only one thing
 at a time, than to understand situations where many things occur
 simultaneously.  Furthermore, we are limited by our main
 communication mechanisms with others, spoken or written language,
 which are inherently sequential.  These convey information in
 parallel through the voice tone, facial and body expressions, writing
 style, etc., but we are often unaware of it.
 Thus, while we are ``parallel processors', and we
 leave in a world where multiple things happen at the same time, we
  usually  reason by reduction to a sequential world.

The same happens in computing. It is much easier to reason about a
sequential program, than about one in which operations are executed
concurrently. 

\paragraph{The grand challenge.}
For more than fifty years, 
one of the most  daunting challenges in information science and
technology lies in mastering concurrency.  Concurrency,
once a specialized discipline for experts, is forcing itself onto the
entire IT community because of two disruptive phenomena: the
development of networking communications, and the end of the ability
to increase processors speed at an exponential rate.  Increases in
performance can only come through concurrency, as in multicore
architectures.  Concurrency is also critical to achieve
fault-tolerant, distributed available services, as in distributed data
bases and cloud computing.
   Yet, software support for these advances lags, mired in concepts
from the 1960s such as semaphores.
The problem is compounded by the inherently non-deterministic nature
of concurrent programs: even minor timing variations may generate
completely different behavior. Sometimes tricky forms of concurrency
faults can appear, such as \emph{data races}
(where even though two concurrent threads handle a
shared data item in a way that is correct from each thread's
perspective, a particular run-time interleaving produces inconsistent
results),  and others, such as
 \emph{deadlocks}
(situation where some threads still have computations to execute,
while none of them can proceed as each is waiting for another one to
proceed), 
 \emph{improper scheduling} of threads or processes, 
 \emph{priority inversions} (where some processes do not proceed
because the resources they need are unduly taken away by others)
and various kinds of \emph{failures} of the processes and
communication mechanisms.  The result is that it is very 
difficult to develop concurrent applications.  Concurrency is
also the means to achieve high performance computing, but in this
paper we are not concerned with such applications.


\paragraph{A classic synchronization difficulty.}
Let us consider the following classical illustration of the concurrency
difficulties in software engineering. A bank account is shared
by a group of people. The rule is that if the balance drops below a
certain threshold, $L$, some high interest will be charged. Thus, each
time a member of the group wants to withdraw some amount of money,
$x$, she first needs to send a message to the bank to make sure the
balance is greater than or equal to $L+x$. Only then she will send a
message asking to withdraw $x$ from the account.  Without any
synchronization, it is impossible to maintain the invariant that the
balance of the account is always at least $L$, unless of course no
withdrawals are ever done. Even assuming that the participants can
directly access the account, synchronization is needed.  Namely,
suppose members of the group can issue $\rread()$ operations, that
directly return the current balance in the account, and execute ${\sf
  withdraw}(x)$ operations that reduce the balance by $x$.  If Alice
asks for the balance and gets back a value $z>L+x$, she cannot then
issue a ${\sf withdraw}(x)$ operation, because she might be a bit
slower than Bob, who could very fast issue $\rread()$ and then ${\sf
  withdraw}(x)$, just after of Alice invoked $\rread()$ but before she
invokes ${\sf withdraw}(y)$.

\paragraph{What does  concurrent computing through sequential thinking mean?}
Instead of trying  to reason directly about concurrent computations,
the idea is to   transform problems in the concurrent domain
into simpler problems in the sequential domain, yielding
benefits for specifying, implementing, and verifying concurrent programs.
It is a two-sided strategy,
together with a bridge connecting them:
\begin{itemize}
\vspace{-0.2cm}
\item Sequential specifications for concurrent programs.
  \vspace{-0.2cm}
\item Concurrent implementations.
  \vspace{-0.2cm}
\item Consistency conditions
   relating concurrent implementations to  sequential specifications.
\end{itemize}

Although a program is concurrent, the specification of the object
(or  service) 
that is implementing is usually through a \textit{sequential specification},
stating  the desired behavior only in executions where the processes
execute one after the other, and often trough familiar paradigms from
sequential computing (such as queues, stacks and lists).  In the
previous example, when we state the rule that if the balance drops
below a certain threshold, $L$, some high interest will be charged, we
are thinking of an account which is always in an {\it atomic} state,
i.e., a state where the balance is well defined.  This makes it easy
to understand the object being implemented, as opposed to a truly
concurrent specification which would be hard or unnatural.  Thus,
instead of trying to modify the well-understood notion of say, a
queue, we stay with the usual sequential specification, and move to
another level of the system the meaning of a concurrent
implementation of a queue.

The second part of the strategy is to provide \textit{implementation
  techniques} for efficient, scalable, and fault-tolerant concurrent
objects.  Additionally, bridging techniques show how to obtain concurrent
executions that appear to the processes as if the operations invoked
on the object where executed atomically, in some sequential
interleaving.  This is captured by the notion of a
\textit{consistency} condition, which defines the way concurrent
invocations to the operations of an object
correspond to a sequential
interleaving, which can then be tested against the its sequential
specification. 

\paragraph{A brief history.}
The history of  concurrency is long;
a few milestones are in the Sidebar~\ref{sidebar:keyDates}.
The interested reader will find many more results in 
      textbooks on shared memory~\cite{L96,R13,T06}, and message-passing 
      systems~\cite{AW04,L96,R18}.  
 We concentrate here only on a few
 significant examples of  sequential reasoning used  to master concurrency,
 highlighting  fundamental  notions of this approach
(such as: sequential specifications, consistency (linearizability),
progress conditions (availability), universal constructions, the need to solve
consensus for fault-tolerance, strong shared objects as a way of
solving consensus, and distributed ledgers). 
We describe several algorithms as a concrete illustration of the ideas.
We tell the story through an evolution that starts with mutual
exclusion, followed by implementing read/write registers on top of message
passing systems, then implementing arbitrary objects, and finally doing so in
a fault-tolerant way through powerful synchronization objects.  We
discuss the modern distributed ledger trends of doing so in a highly scalable,
tempered-proof way.

We conclude with a discussion of the limitations of this approach: it
is expensive to require atomicity (linearizability), and furthermore,
there are inherently concurrent problems with no sequential specifications.

\begin{sidebar}[ht!]
\fbox{
\centering{
\begin{minipage}[t]{150mm}
      {\scalebox{0.43}[0.43]{\input{fig-history-dates-long-V2.pstex_t}}}
      ~\\~\\ $~~~~~~~~$  
 Some of the previous papers were
     awarded the famous ACM-EATCS Dijkstra Award. Created in 2000, this
      award is given to outstanding papers on the principles of
      distributed computing, whose significance and impact on the
      theory and/or practice of distributed computing have been
      evident for at least a decade). In the history line,
      ``[aa] DA bcde'' means  ``paper(s) referenced [aa]
      received the  Dijkstra Award in the year bcde''. 

 The interested reader will find recent
      textbooks on shared memory synchronization
      in~\cite{HS08,R13,HS08,T06}, and message-passing synchronization
      in~\cite{AW04,CGR11,G02,KS08,L96,R13,R18,S07}. 
\end{minipage}
}
\caption{History of synchronization: a few important  dates}
\label{sidebar:keyDates}
}
\end{sidebar}

\newpage

\section*{MUTUAL EXCLUSION}
Concurrent computing began in 1961 with what was called
\textit{multiprogramming} in the Atlas computer, where concurrency was
simulated --~as we do when telling stories where things happen
concurrently~-- interlacing the execution of sequential programs.
Concurrency was born in order to make efficient use of a sequential
computer, which can execute only one instruction at a time, giving
users the illusion that their programs are all running simultaneously,
through the operating system.  A collection of early 
foundational articles on concurrent programming appears in~\cite{B02}.

As soon as the programs being run concurrently began to interact with
each other, it was realized how difficult it is to think concurrently.
By the end of the 1960s there was already talk of a crisis:
programming was done without any conceptual foundation and lots of
programs were riddled with subtle errors  causing erratic
behaviors. In 1965 Dijkstra discovered that the \textit{mutual
  exclusion} of parts of code is a fundamental concept of programming,
and opened the way for the first books of principles on concurrent
programming which appeared at the beginning of the 1970s.

1970s, multi-processor computers were built, and in 1967 there were
debates about their computing computing and what is known today as
Amdahl's Law. In the late 1970s a move occurred from multiprocessors
with shared memory to multicomputers with distributed memory
communicating by sending messages. In the 1990s, the importance of
shared memory with multicores returns, as it meets the barriers of
energy expenditure, and the limits of making processors increasingly
faster, emphasizing that the exponential growth prophesied by Moore's
Law refers to packaging more and more components on the same chip,
that is more and more parallel computing. And now entering a
new generation of distributed systems motivated by new distributed
services and blockchain-like applications, where we are building
huge distributed systems open and tolerant to arbitrarily malicious faults.
Nevertheless, while both parallel computing and distributed computing
involves concurrency, they are different address different
computing worlds (see Sidebar~\ref{sidebar-terminology}).

\begin{sidebar}[h!]
\fbox{
\centering{
  \begin{minipage}[t]{150mm}
   $~$\\
  As far as terminology is concerned we consider the following definitions
  (from~\cite{R18}).
\begin{itemize}
\item  {\it Parallel computing}.
  Parallel computing addresses concepts, methods, and strategies which
  allow us to benefit from parallelism (simultaneous execution of distinct
  threads or processes) when one has to implement a computation.
  The essence of parallel computing lies in the decomposition of the
  computation in independent computation units and exploit their independence
  to execute as many of them as possible in parallel (simultaneously) so
  that the resulting execution is time-efficient.
\vspace{-0.2cm}
\item {\it Distributed computing}.  Distributed
  computing arises when one has to solve a problem involving
  geographically distributed entities (processors, nodes, sensors,
  peers, agents, etc.), such that each entity only has a partial
  knowledge of the many input parameters involved in the problem to be
  solved. Because their knowledge is partial, these computing entities
  must cooperate to solve the problem. They also must cope with their
  environment, which can be modeled as adversaries, such as
  asynchrony, failures, mobility, etc. These adversaries create an
  uncertainty on the state of the system, uncertainty that has to be
  understood and mastered if one wants to produce correct distributed
  software. 
\end{itemize}

  As we can see, parallel and distributed computing are in some sense dual:
  one consists in decomposing a computation into independent entities,
  while the other   consists in allowing pre-existing entities
  --~whose distribution is not under the control of the programmer~--
  to cooperate  in the presence of adversaries such as the net effect of
  asynchrony and process  failures.
\end{minipage}
}
\caption{Distributed computing versus  parallel computing:
  the two faces of concurrency}
\label{sidebar-terminology}
}
\end{sidebar}

\paragraph{Mutual exclusion.}
A mutual exclusion algorithm consists of the code for two operations,
${\sf acquire}()$ and ${\sf release}()$, that a process invokes to
bracket a section of code called a {\it critical section}.  The usual
environment in which a mutual exclusion algorithm is executed is {\it
  asynchronous}, where  process speeds are arbitrary,
independent from each other.  The mutual exclusion algorithm should
guarantee two conditions.
\begin{itemize}
  \vspace{-0.2cm}
\item Mutual exclusion. No two processes are
simultaneously executing their critical section.
\vspace{-0.2cm}
\item Deadlock-freedom.  if
one or several processes invoke concurrently ${\sf acquire}()$,
eventually one of them terminates its invocation, and consequently
executes its critical section.
\end{itemize}

\paragraph{Progress conditions.}
When Dijkstra introduced mutual exclusion~\cite{D65}, he also
introduced the previous progress condition, called {\it
  deadlock-freedom}.  As observed by D.E. Knuth in~\cite{K66},
Deadlock-freedom does not prevent specific timing scenarios from
occurring in which some processes can never enter their critical
section.  Hence, he proposed The stronger {\it starvation-freedom}
progress condition, states that any process that invokes ${\sf
  acquire}()$ will terminate its invocation (and will consequently
execute its critical section).

\paragraph{On mutual exclusion algorithms from atomic read/write registers.}
The first mutual exclusion algorithms were abstruse, difficult to
understand and prove correct (some of them are collected in~\cite{R87}). 
 We describe here an elegant algorithm by
Peterson~\cite{P81}.  The version presented in
Algorithm~\ref{fig:Peterson-2-processes} is for two processes, but can
be easily generalized to $n$ processes.

The two processes $p_1$ and $p_2$ share three read/write atomic
registers, $\mathit{FLAG}[1]$, $\mathit{FLAG}[2]$, and
$\mathit{LAST}$.  Initially $\mathit{FLAG}[1]$, $\mathit{FLAG}[2]$,
are down, while $\mathit{LAST}$ does not need to be initialized.  Both
processes can read all registers. Moreover, while $\mathit{LAST}$ can
be written by both processes, only $p_i$, $i\in\{1,2\}$, writes to
 $\mathit{FLAG}[i]$.
{\it Atomic} means that the read and write operations on the registers
appear as if they have been executed sequentially (hence,
the notion of ``last writer'' associated with  $\mathit{LAST}$ is well defined).

\begin{algorithm}[h]
\centering{
\fbox{
\begin{minipage}[t]{150mm}
\footnotesize
\renewcommand{\baselinestretch}{2.5}
\resetline
\begin{tabbing}
aA\=aA\=aaA\=aaA\=aaA\=aaA\kill
{\bf operation} ${\sf acquire}()$ {\bf is}   $~~~$ \% invoked by $p_i$,
                                                      $i\in \{1,2\}$ \\
\>    $\mathit{FLAG}[i] \leftarrow {\tt up}$;
      $\mathit{LAST} \leftarrow i$;   {\bf let} $j=3-i$;\\
\>     {\bf wait} \big($(\mathit{FLAG}[j]= {\tt down})~ \vee~
                        (\mathit{LAST} \neq i)$\big); \\
\> ${\sf return} ()$\\
{\bf end operation}.
\\~\\
{\bf operation} ${\sf release}()$ {\bf is} 
  $\mathit{FLAG}[i] \leftarrow  {\tt down}$; 
  ${\sf return} ()$ 
{\bf end operation}.

\end{tabbing}
\normalsize
\end{minipage}
}
\caption{Peterson's algorithm for two processes}
\label{fig:Peterson-2-processes}
}
\end{algorithm}

When process $p_i$ invokes ${\sf acquire}()$, it first raises its flag,
thereby indicating it is competing,   and then writes its name in
$\mathit{LAST}$ indicating it is the last writer of this register. 
Next process $p_i$ repeatedly reads $\mathit{FLAG}[j]$ and
$\mathit{LAST}$ until it sees $\mathit{FLAG}[j]={\tt down}$ or it is
no longer the last writer of $\mathit{LAST}$.  When this occurs, $p_i$
terminates its invocation.  The operation ${\sf release}()$ consists
in a simple lowering of the flag of the invoking process.
 The read and write operations
on $\mathit{FLAG}[1]$, $\mathit{FLAG}[2]$, and $\mathit{LAST}$ are totally
ordered (atomicity), which  facilitates  the proof of the
 mutual exclusion and starvation-freedom properties. \\

 Mutual exclusion was the first mechanism for mastering concurrent
 programming through sequential thinking, and lead to the
 identification of notions that began to give a scientific foundation
 to the approach, such as the concepts of \textit{progress condition} and
 \textit{atomicity}.

 \paragraph{Fast mutual exclusion and adaptive algorithms.}
 The previous algorithm can be easily generalized to solve mutual
exclusion in a set of $n\geq 2$ processed.  Many $n$-process mutual
exclusion algorithms have been proposed, in which each process must
solve $(n-1)$ conflicts to access the critical section.  An algorithm
in which the number of read and write accesses to shared registers is
constant in contention-free scenarios appears in Lamport~\cite{L87}.
This article is the origin of research on {\it adaptive algorithms},
whose complexity depends on the concurrency pattern in which
operations are invoked.

\paragraph{Atomicity from non-atomic read/write registers.}
The previous algorithms implements mutual
exclusion using underlying atomic read/write registers.  In fact,
this hardware atomicity is not required, Lamport~\cite{L74}, showed
that mutual exclusion can be achieved using only {\it safe}
registers~\cite{L86}. Several algorithms building atomic read/write
registers from non-atomic read/write registers are described in
e.g.,~{\cite{R13,T06}.

\paragraph{On the database side.}
The concept of a {\it transaction} was introduced in database 
as a computation unit (usually, an operation-based translation
of a query expressed in a specific query language)~\cite{GR92}.
The management of transactions introduced the notion of
{\it concurrency control}, which gave rise to several approaches
to ensure that transactions appear as if they had been executed
sequentially~\cite{BHG87,P79}.

\paragraph{Transactional memory.}
The concept of {\it transactional memory} (TM) was introduced by
M. Herlihy and J. Moss in 1993~\cite{HM93}, and then investigated from
a pure software point of view (STM) by N. Shavit and D. Touitou in
1997~\cite{ST97}.

The aim of a TM/STM system is to discharge the programmers from the
management of synchronization in multiprocess programs that access
concurrent objects. To that end, an TM/STM system provides the
programmer with the concept of a transaction.  Basically, the job of
the programmer is to design each process of the application as a
sequence of transactional code and non-transactional code, where a
transaction is any piece of code that accesses concurrent objects, but
contains no explicit synchronization statement, and non-transactional
code does not access concurrent objects.  It is then the job of the
underlying TM/STM system to provide the illusion that each transaction
appears as being executed atomically (see
Sidebar~\ref{sidebar:atomic-register}, where each read or write
operation is replaced by a transaction). Executing each transaction in
a critical section would solve the problem, but this would be
inefficient. So, for efficiency, a TM/STM system must allow
transactions to execute concurrently. The major parts of a TM/STM
systems execute transaction in a speculative mode at the end of with a
transaction is committed or aborted.  According to the TM/STM
system, the recovery of a transaction can be under the control of
either the system  or the invoking process.  Examples of STM systems based on
different underlying principles can be found in~\cite{CR06,DSS06}.

As we can see, a TM/STM system allows the programmer to concentrate
on the problem it has to solve and not on the way the required
synchronization must be  implemented. In this sense it provides
the programmer with a higher abstraction level. 
It is important to see that a transaction canbe any piece of code
(and not a code obtained from a specific query langauge as in databases). 
TM/STM provides programmers with a tool from which they can see
executions as sequences of transactional codes. \\

The important point here is that both concurrency control in database and
transactional memory aim at providing an abstraction level at which
the users see an execution as if it was produced by a sequential processor.

\section*{FROM RESOURCES TO OBJECTS}

\paragraph{From physical resources to services.}
At the beginning, a critical section was encapsulating the use of a
physical resource, which by its own nature, is sequentially specified
(e.g., disk, printer, processor).  Conceptually not very different, it
was then used to protect concurrent accesses to preserve consistency
of simple data (such as a file in the readers/writers problem~\cite{CHP71}).
However, when critical sections began to be used to encapsulate more
general {\it shared objects}, new ideas were needed.

\paragraph{Data are not physical resources.} 
 A shared object is different from a physical object, in that it does
 not a priori require exclusive access; a process can read the data of
 a file while another process concurrently modifies it.
 The {\it mutex-free} (also called {\it lock-free})
 approach  (introduced by Lamport
 in~\cite{L77}), makes possible to envisage implementations of purely
 digital objects in which operation executions
are free from mutual exclusion and can overlap in time,
none of them depending of the others to terminate~\cite{H91}
(see progress conditions defined below).

\paragraph{Consistency conditions.}
Wherever concurrent accesses to share data take place, a consistency
condition is needed to define what does it mean to correctly execute
concurrently operations, especially in the presence of buffers and
memory caches (that are defined only in sequential executions, such as
read/write operations).  Instead of transforming a concurrent
execution into sequential execution (as in mutual exclusion), the idea
appears to enforce only {\it virtual sequentiality}, namely, from an
external observer point of view, everything must appear as if the
operations were executed sequentially, thereby reducing --at a higher
abstraction layer-- concurrent computing to sequential computing.
When the total order on the operations is required to respect the
order on non-overlapping operations, this virtual sequentiality is
called {\it atomicity}~\cite{L86} or {\it linearizability}~\cite{HW90}
(these two terms are synonyms).  This is illustrated in
Sidebar~\ref{sidebar:atomic-register}, which describes an execution in
which three processes access an atomic read/write register $R$.

\paragraph{A short historical perspective.} 
Since early on in 1976, in the database context,
\textit{serializability}~\cite{P79,SLR76} of transactions that aggregate
many operations without locking and unlocking entities was generally
accepted as the right notion of correctness: to require that
transactions appear to have executed atomically.  In the concurrent
programming the equivalent notion of \textit{sequential consistency}
was used, but for individual operations~\cite{L79}.  Later on,
realizing that this type of condition is not composable,
\textit{linearizability}~\cite{HW90} required additionally that this
sequential order must also preserve the global ordering of
non-overlapping operations.  Linearizability may be preferred over
sequential consistency, because a system made of linearizable
implementations is linearizable.

\paragraph{The environment has an impact on computations: crash failures.}
Let us remark  that mutual exclusion cannot work
when one has to implement an object  in the presence of asynchrony
and process  \textit{crashes} (premature halting). 
If a process crashes inside its critical section,  mutual exclusion will
never be released, and no other process will be able to access the object.
It follows that the use of mutual exclusion (locks)
is limited in the presence of asynchrony and process crashes.

\begin{sidebar}[ht]
\fbox{
\centering{
\begin{minipage}[]{150mm}
\centering{\scalebox{0.45}{\input{fig-registre-atomique-exemple.pstex_t}}}
~\\~\\~An atomic (linearizable) execution of processes $p_1,p_2$, and
$p_3$ on atomic register $R$.  The read and write operations are
denoted $R.{\sf read}()$ and $R.{\sf write}()$.  From an external
observer point of view, it appears as if the operations were executed
sequentially.
\end{minipage}
}
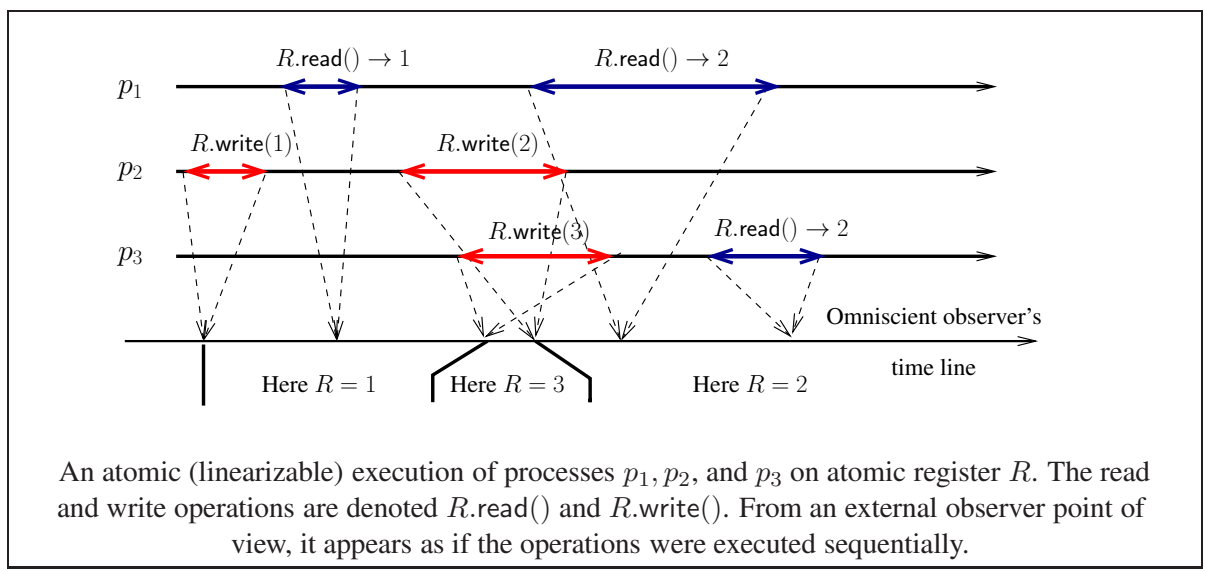
\caption{An atomic execution of a read/write register}
\label{sidebar:atomic-register}
}
\end{sidebar}

\paragraph{On progress conditions in the presence of crash failures.} 
Three progress conditions have been proposed
for the implementation of the operations of  data objects
in an environment where processes are asynchronous and  may crash.
They are the following ones, going from the stronger to the weaker
(see Table~\ref{table:progress-conditions}). 
\begin{itemize}
\vspace{-0.2cm}
\item    
The {\it wait-freedom} progress condition states that
if a process invokes an object operation, and does not crash,
it terminates its invocation~\cite{H91}. This means that it terminates
whatever the behavior of the other processes (e.g., some of them being
crashed, and others being concurrently involved in object operations). 
\vspace{-0.2cm}
\item 
The {\it non-blocking} progress condition states that
if several processes concurrently invoke operations on the object,
at least one of them terminates~\cite{HW90}.
\vspace{-0.2cm}
\item 
The {\it obstruction-freedom} progress condition states that
if a process invokes an operation, does not crash during this invocation,
and all other processes stop accessing the internal representation of
the object  during a long enough period, then the process terminates its
operation~\cite{HLM03}.
\end{itemize}
  
Let us remark that the wait-freedom and non-blocking progress conditions
are independent of both the failure pattern  and the concurrency pattern.
They can be seen as the ``corresponding'' of starvation-freedom and
deadlock-freedom in asynchronous crash-prone system.
Differently, obstruction-freedom is dependent on the concurrency pattern.

\begin{table}[htbp]
\begin{center}
\renewcommand{\baselinestretch}{1} \small
\begin{tabular}{|c|c|}
\hline         Lock-based implementations & Mutex-free implementations    \\   
\hline    
\hline                             & Obstruction-freedom  ~\cite{H91}  \\
\hline         Deadlock-freedom ~\cite{D65}   & Non-blocking   ~\cite{HW90} \\
\hline         Starvation-freedom ~\cite{K66}  & Wait-freedom   ~\cite{HLM03} \\
\hline  
\end{tabular}
\end{center}
\vspace{-0.3cm}
\caption{Progress conditions for the implementation of concurrent objects}  
\label{table:progress-conditions} 
\end{table}

\section*{READ/WRITE REGISTERS ON TOP OF MESSAGE-PASSING SYSTEMS}
The read/write shared register abstraction provides several
advantages over  message passing: a more
natural transition from uniprocessors, and  simplifies 
programming tasks.
For this reason, concurrent systems that support shared memory are
have wide acceptance in both research and commercial computing.
  
It is relatively easy to build atomic read/write registers on top of a
reliable asynchronous message-passing system (e.g.~\cite{R13-MP}),
but if processes may crash, more
involved algorithms are needed.  Two important results are
presented by Attiya, Bar-Noy and Dolev in~\cite{ABD95}:
\begin{itemize}
\vspace{-0.2cm}
\item An algorithm that implements an atomic read/write register on
  top of a system of $n$ asynchronous message-passing processes, where
  at most $t<n/2$ of them may crash.
  \vspace{-0.2cm}
\item A proof of the impossibility of building an atomic read/write
  register when $t\geq n/2$.
\end{itemize}
The section presents the algorithm, referred to as the \textit{ABD
  Algorithm}, which illustrates the importance of the ideas of reducing
concurrent thinking to sequential reasoning. A more detailed proof can
be found in~\cite{ABD95,AW04,R18}, as well as other algorithms.

\paragraph{Design principles of ABD: each written value has an identity.}
Each process  is both a client and a server.
Let $\REG$ be the multi-writer multi-reader (MWMR) register that is built
(hence any process is allowed to  read and write the register). 
On its client side a process $p_i$ can invoke the operations
$\REG.\wwrite~(v)$ (to write a value $v$ in $\REG$, and 
$\REG.{\sf read}~()$ to obtain its current value. 
On its server side, a process $p_i$ manages two local variables:
$reg_i$ which locally implement $\REG$, and $timestamp_i$
which  contains a timestamp made up of a sequence number
(which can be considered as a date) and a process identity $j$. 
The timestamp $timestamp_i$ constitutes the ``identity'' of the value $v$
saved in $reg_i$ (namely, this value was written by this process at
this time).  Any two timestamps $\langle sn_i,i\rangle$ and
$\langle sn_j,j\rangle$ are totally ordered by their lexicographical order;
namely, $\langle sn_i,i \rangle < \langle sn_j,j\rangle$ means
$(sn_i < sn_j) \vee (sn_i=sn_j \wedge i<j)$.

\paragraph{Design principles of ABD: intersecting quorums.}
The basic mechanism on which ABD relies on a
query/response message exchange pattern.  A process $p_i$ broadcasts a
query to all the processes and waits for acknowledgments from a
majority of them. Such a majority \textit{quorum} set, has the
following properties.  As $t<n/2$,
waiting for acknowledgments from a
majority of processes can never block forever the invoking process.
Moreover, the fact that any two quorums have a non-empty intersection
implies the atomicity property of the read/write register $\REG$.

\begin{algorithm}[h]
\centering{
\fbox{
\begin{minipage}[t]{80mm}
\footnotesize
\renewcommand{\baselinestretch}{2.5}
\resetline
\begin{tabbing}
aaaaA\=aaaaaaaaaA\=aaaaaaaaaaaaA\kill

{\bf operation} $\REG.\wwrite~(v)$ {\bf issued by process} $p_i$  {\bf is} \\

\>  build a new tag  $tag$ identifying this write operation;\\

\> \% Phase 1: acquire information on the system state \%  \\
 
\>  $\bbroadcast$ {\sc write\_req} $(tag)$;\\
 
\> $\wwait$ acknowledgments  from a majority of processes,\\
\>  each  carrying $tag$ and a sequence number;\\

\> \% Phase 2 : update  system state \%  \\

\>  $ts \leftarrow \langle msn+1,i \rangle$ where
    $msn$ is \\
\>    the greatest  sequence number  previously received; \\

\> $\bbroadcast$  {\sc write} $(tag,v,ts)$;\\

\> $\wwait$ acknowledgments carrying $tag$ from a majority of proc.;\\

\>  ${\sf return} ()$.\\~\\

{\bf when} {\sc write\_req} $(tag)$ {\bf is received from}  $p_j$, 
$j\in\{1,...,n\}$ {\bf do}\\
                                    
\>  $\ssend$ to $p_j$ an  acknowledgment carrying $tag$,
        and \\
\>        the sequence number contained in  $timestamp_i$.\\~\\

 {\bf when} {\sc write} $(tag,v,ts)$  
               {\bf is  received from} $p_j$, $j\in\{1,...,n\}$  {\bf do}\\

\> {\bf if}  \=  $(timestamp_i< ts)$ 

{\bf then}  \\
\>\> $timestamp_i  \leftarrow  ts$;
            $reg_i \leftarrow v$  {\bf end if};\\

\> $\ssend$ to $p_j$  an  acknowledgment carrying $tag$.

\end{tabbing}
\normalsize
\end{minipage}
}
\caption{Operation  $\REG.\wwrite~(v)$:
client and server behavior for a process $p_i$}
\label{fig-MWMR-atomic-register-write}
}
\end{algorithm}

\paragraph{The operation   $\REG.\wwrite~(v)$.}
This operation is implemented by Algorithm
\ref{fig-MWMR-atomic-register-write}. 
When a process $p_i$  invokes $\REG.\wwrite~(v)$, it first 
creates a tag denoted ($tag$) which will identify the 
query/response messages generated by this write invocation.
Then (phase 1), it executes a first instance of the query/response exchange
pattern to learn the highest sequence number saved in the local variables
$timestamp_j$ of a majority of processes $p_j$.
When this is done, $p_i$ computes the timestamp $ts$ which will be
associated with the value $v$ it wants to write in $\REG$. 
Finally (phase 2), $p_i$ starts a second query/response pattern
in which it broadcasts the pair $(v,ts)$ to all the processes.
When, it has received the associated acknowledgments from a quorum, 
$p_i$ terminates the write operation. 

On its server side, a process $p_i$ that receives a {\sc write\_req}
message sent by a process $p_j$ during phase 1 of a write operation,
sends it back an acknowledgment carrying the sequence number
associated with the last value it saved in $reg_i$.  When it receives
{\sc write\_req} message sent by a a process $p_j$ during phase 2 of a
write operation, it updates its local data $reg_i$ implementing $\REG$
if the received timestamp is more recent (with respect to the total
order on timestamps) than the one saved in $timestamp_i$ , and, in all
cases, it sends back to $p_j$ and acknowledgment (so $p_j$ terminates
its write).

It is easy to see that, due to the intersection property of 
quorums, the timestamp  associated with a value $v$ by the invoking process
$p_i$ is greater than the ones of the write operations that terminated
before $p_i$ issued its own write operation.  Moreover, while
concurrent write operations can associate the same sequence number
with their values, these values have different (and ordered)
timestamps.

\paragraph{The operation   $\REG.\rread~()$.}
Algorithm~\ref{fig-MWMR-atomic-register-read} implements 
operation operation   $\REG.\rread~()$, with a similar structure as
the implementation of operation $\REG.\wwrite~()$.
Namely, it is made up of
two phases, each one being an instance of the query/response
communication pattern.  In the first phase, the invoking process
obtains a pair (value, associated timestamp) from a minority of
processes, from which --~thanks to the total order on timestamps~-- it
can extract the most recent value, that it will return as the result
of the read operation.

\begin{algorithm}[h]
\centering{
\fbox{
\begin{minipage}[t]{150mm}
\footnotesize
\renewcommand{\baselinestretch}{2.5}
\resetline
\begin{tabbing}
aaaaA\=aaaaaaaaaA\=aaaaaaaaaaaaA\kill
{\bf operation}  $\REG.{\sf read}~()$ {\bf is}\\

\>  build a new tag  $tag$ identifying this read operation;\\
\> \% Phase 1: acquire information on the system state \%  \\
 
\>  $\bbroadcast$ {\sc read\_req} $(tag)$;\\
 
\> $\wwait$
acknowledgments from a majority of processes, \\
\> each carrying $tag$ and a pair $\langle$value,timestamp$\rangle$;\\

\>{\bf let} $ts$  be the greatest timestamp received, \\ 

\> and  $v$ the value associated with this timestamp;\\

\> \% Phase 2 : update  system state \%  \\

\> $\bbroadcast$ {\sc write} $(tag,v,ts)$;\\

\> $\wwait$ {\sc ack\_write} $(tag)$ from a majority of proc.;\\
\> $\return~(v)$.\\~\\


{\bf when} {\sc read\_req} $(tag)$ {\bf is received from}
$p_j$,  $j\in\{1,...,n\}$  {\bf do}
\\

\>  $\ssend$  to $p_j$  an  ack. carrying $tag$, $reg_i$ and $timestamp_i$. 
  
\end{tabbing}
\normalsize
\end{minipage}
}
\caption{Operation  $\REG.\rread~()$:
client and server behavior for a process $p_i$}
\label{fig-MWMR-atomic-register-read}
}
\end{algorithm}

Notice that the following scenario can occur,
which involves  two read operations $\rread1$ and $\rread2$ on a register
$\REG$ by the  processes $p_1$ and $p_2$, respectively, and a  
concurrent write operation $\REG.\wwrite(v)$ issued by a process $p_3$
(Fig.~\ref{fig:new-old-inversion}).
Let $ts(v)$ be the timestamp associated with $v$ by $p_3$.

\begin{figure}[h!]
\centering{
\begin{minipage}[]{150mm}
\centering{\scalebox{0.35}{\input{fig-new-old-inversion.pstex_t}}}
\end{minipage}
}
\caption{New/old inversion scenario}
\label{fig:new-old-inversion}
\end{figure}
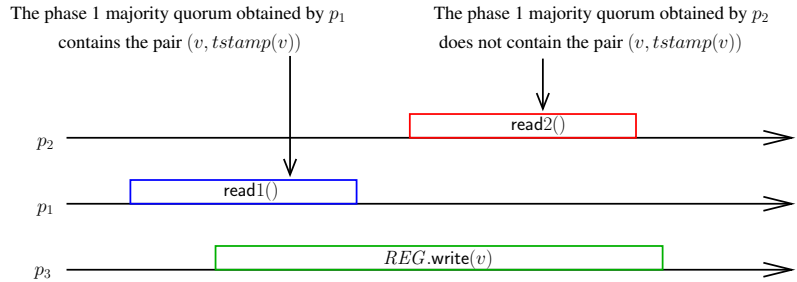


It is possible that the phase 1 majority quorum obtained by $p_1$
includes the pair ($v,ts(v)$), while the one obtained by $p_2$
does not.  If this occurs, the first read operation $\rread1$ obtains
a value more recent that the one obtained by the second $\rread2$,
which violates atomicity. This can be easily solved by directing each
read operation to write the value it is about to return as a
result. In this way, when $\rread1$ terminates and returns $v$, this
value is known by a majority of processes despite asynchrony,
concurrency, and a minority of process crashes. This phenomenon
(called {\it new/old inversion}) is prevented by the phase 2 of a read
operation.

The combination of intersecting quorums and timestamps
allows for the implementation of atomic read/write registers in
asynchronous message-passing systems where a minority of process may
crash.  Hence, sequential thinking on shared registers can be used at
the upper abstraction level.

\section*{THE WORLD OF CONCURRENT OBJECTS}
\paragraph{Objects defined by a sequential specification.}
A read/write register is a special case of an immaterial
{object}.
In general, an \textit{object} is defined by the set of operations that
processes can invoke, and by an automaton, which specifies the
behavior of the object when these operations are invoked sequentially.
The automaton specifies, for each state, and each possible operation
invocation, a response to that invocation, and a transition to a new
state.  A stack for example, is easily specified in this way.  The
operations are ${\sf push}(v)$, to add $v$ at the top of the stack; if
the stack is full, it returns the control value ${\tt full}$. Similarly, if
the stack is not empty, the operation ${\sf pop}()$ returns the value
at the top of the stack and suppresses it from the stack; and it returns
the control  value ${\tt empty}$ if the stack is empty.

A concurrent stack can be implemented by executing  the operations
${\sf pop}()$  and  ${\sf push}()$ using   mutual exclusion.
As already indicated, this  strategy to create
a total order does not work if processes  may crash. 
The {\it state machine replication} mechanism~\cite{L78}
is a general way of implementing  an object by asynchronous
crash-prone processes, that  invoke operations on the object concurrently.

Implementing a state machine is easy if no process crash.
This is no longer the case in crash-prone asynchronous systems, where
the implementation of a state machine relies on the consensus object.

\paragraph{Consensus.}
At the core of many sequential reasoning for concurrent programming
situations (including state machine replication) are agreement problems.
A common underlying abstraction  is the  \textit{consensus} object.
It has  a single operation
denoted ${\sf propose}()$, that a process can invoke once. 
If a process invokes  ${\sf propose}(v)$, the
 invocation  eventually returns a value $v'$.
 This sequential specification is defined by the following properties.
\begin{itemize}
\vspace{-0.2cm}
\item Validity. If an invocation returns $v$ then there is a
  ${\sf propose}(v)$.
\vspace{-0.2cm}
\item Agreement.  No two  different values are returned.
  \vspace{-0.2cm}
\item Termination. If a process invokes  ${\sf propose}()$ and does not
  crash, it returns a value.  
\end{itemize}
Consensus objects are universal in the sense that (together with
read/write registers), they can be used to implement, despite asynchrony and
process crashes, any  object defined by a sequential
specification.  The consensus-based state machine replication technique
provides an illustration  of this claim, as discussed below.

\paragraph{All objects are not equal  in a crash-prone environment.}
It turns out that an object as simple as a concurrent stack cannot be
implemented by asynchronous processes, which communicate using
read/write registers only, if any operation invoked by a process that
does not crash must return (independently of the speed or crashes of
the other processes).  Such an implementation of an object is said to
be \textit{wait-free}~\cite{H91}.

A way of measuring the synchronization power of an object in the
presence of asynchrony and process crashes is by its {\it consensus
  number}~\cite{H91}.  The consensus number of an object $O$ is the
greatest integer $n$, such that it is possible to wait-free implement
a consensus object for $n$ processes from any number of objects $O$
and atomic read/write registers. The consensus number of  $O$
is $\infty$ is there is no such greatest integer.
As an example, the consensus number of a Test\&Set object or a stack object
is $2$, while  consensus number of a Compare\&Swap or LL/SC object is $\infty$. 
The power and limits of  shared memory systems is addressed in~\cite{HRR13}.


\section*{STATE MACHINE REPLICATION}
 The {\it state machine replication} mechanism~\cite{L78} is the main
 approach to implement an object in a concurrent system, with
 asynchrony and process crash failures in message-passing
 systems~\cite{L78,S90}, and in multiprocessors where each processor has
 a local memory~\cite{R13}.  The idea is for the processes to agree on
 a sequential order of the concurrent invocations, and then  each
 one to simulate the sequential specification automaton locally. 
 We illustrate here the approach with a mechanism for
 reaching the required agreement: a total order broadcast abstraction.

\paragraph{Total order broadcast.}
The TO-broadcast abstraction~\cite{HT94,R18} in an important primitive
in distributed computing, that ensures that all correct processes  receive
 messages in the  same order (we do not define them more formally here).
It is used through two operations, ${\sf TO\_broadcast}()$ and ${\sf
  TO\_deliver}()$.  A process invokes ${\sf TO\_broadcast}(m)$, to
send a message $m$ to all other processes.  As a result, processes
execute ${\sf TO\_deliver}()$ when they receive a (totally ordered) message.
The TO-broadcast abstraction is defined by the following properties
(the first three are safety, while the last two are liveness
properties).  It is assumed without loss of generality that all
messages are different.
\begin{itemize}
  \vspace{-0.2cm}
\item TO-validity. If a
  process  executes ${\sf TO\_deliver}(m)$ (i.e., to-delivers the a message $m$)
  , then a process
executes ${\sf TO\_broadcast}(m)$.
\vspace{-0.2cm}
\item TO-integrity.  If a process
executes ${\sf TO\_deliver}(m)$ and ${\sf TO\_deliver}(m')$, then
$m\neq m'$.
\vspace{-0.2cm}
\item TO-order. If a process executes ${\sf
TO\_deliver}(m)$ and ${\sf TO\_deliver}(m')$ in this order, then no
process executes these operations in the reverse order.
\vspace{-0.2cm}
\item
TO-termination-1.  If a process executes ${\sf TO\_broadcast}(m)$ and
does not crash, it eventually  executes ${\sf TO\_deliver}(m)$.
\vspace{-0.2cm}
\item
TO-termination-2. If a process executes ${\sf TO\_deliver}(m)$, then
every process that does not crash executes ${\sf
  TO\_deliver}(m)$.
\end{itemize}

 TO-broadcast illustrates one more general idea within the theory of
 mastering concurrent programming through sequential thinking: the
 identification of communication abstractions that facilitate building
 concurrent objects defined by a sequential specification.

\paragraph{State machine replication based on TO-broadcast.}
A concurrent implementation of object $O$ is described in
Algorithm~\ref{fig:TO-URB-based-universal-construction}.  
It is a {universal construction}, as it
works for any object $O$ defined by a sequential specification.  The
object has operations ${\sf op}_x()$, and a transition function
$\delta()$ (assuming $\delta$ is deterministic), where
$\delta(state,{\sf op}_x(param_x))$ returns the pair $\langle
state',r\rangle$, where $state'$ is the new state of the object and
$res$ the result of the operation.

\begin{algorithm}[h]
\centering{
\fbox{
\begin{minipage}[t]{150mm}
\footnotesize
\renewcommand{\baselinestretch}{2.5}
\resetline
\begin{tabbing}
aaaaA\=aaaA\=aaaaaaaaaaaaA\kill

{\bf when operation} ${\sf op}_x ~(param_x)$ 
     {\bf is invoked by}  the client $p_i$ {\bf do}\\

\>   $result_i  \leftarrow  \bot$; 

  {\bf let} $sent\_msg = \tup{{\sf op}_x ~(param_x),i}$;\\

\>   ${\sf TO\_broadcast}~( sent\_msg)$; \\
      
\>   $\wwait$ $(result_i\neq\bot)$; 

${\sf return} ~(result_i)$.\\~\\


{\bf background task} $T$ {\bf is}\\

\> {\bf repeat forever}\\

\> \>  $rec\_msg \leftarrow {\sf TO\_deliver}()$;\\ 

\> \>  $\langle state_i,res\rangle \leftarrow \delta(state_i,rec\_msg.op)$;\\ 

\> \>  {\bf if} $(rec\_msg.\mathit{proc} =i)$  {\bf then} 
                 $result_i \leftarrow res$  {\bf end if}\\

\> {\bf end repeat}.

\end{tabbing}
\normalsize
\end{minipage}
}
\caption{TO-broadcast-based construction}
\label{fig:TO-URB-based-universal-construction}
}
\end{algorithm}

Let $p_1,$, ..., $p_n$ be the set
of asynchronous crash-prone processes.  Each process $p_i$ is both
client (it can invoke operations on $O$) and server (it participates
in the implementation of $O$).
The idea of the construction is simple.
Each process $p_i$ has a copy $state_i$ of
the object, and the TO-broadcast abstraction is used to ensure that
all the processes $p_i$ apply the same sequence of operations to their local
representation $state_i$ of the object $O$.  
When a process $p_i$ invokes an operation it builds a message
$sent\_msg$ composed of two fields: $sent\_msg.op$  which contains
the operation and  $sent\_msg.proc$ which contains the identity of
the invoking process. Then $p_i$ to-broadcasts  $sent\_msg$ 
and waits until its operation has been executed on its local copy of $O$.
%
On it server side, a process $p_i$ executes an infinite loop in which it
first waits for the next message to-delivery. 
Then, it computes the next state of the object $O$, and,
if it is the process that invoked the operation, it
writes its result into  its local variable $result_i$ to allow the
operation to terminate. 
The correction of this simple universal construction follows directly from
the properties of the to-broadcast abstraction~\cite{HT94,R18}.

\paragraph{Implementing TO-broadcast from consensus.}
Algorithm~\ref{fig:algo-tobroadcast-from-consensus} is a simple construction
of   TO-broadcast on top of an asynchronous system enriched with
consensus objects~\cite{HT94}.

Each process $p_i$ manages four local variables:
a sequence number $sn_i$ initialized to $0$,  a set of message $delivered_i$
initialized to $\emptyset$, a queue $to\_deliverable_i$ initialized to
the empty sequence $\epsilon$, and an auxiliary variable $res_i$.
Let $\bbroadcast(m)$  stand for
``{\bf for each} $ j\in\{1,...,n\}$ {\bf do}
$\ssend(m)$ $\tto$ $p_j$  {\bf end for}''.  If the invoking process
does not crash during its invocation, all processes
receive $m$; if it crashes an arbitrary subset of processes receive $m$. 
To simplify the presentation, it is assumed that a process can send a
message to itself.

\
\begin{algorithm}[h]
\centering{
\fbox{
\begin{minipage}[t]{150mm}
\footnotesize
\renewcommand{\baselinestretch}{2.5}
\resetline
\begin{tabbing}
aaA\=aA\=aaaaaaaaaaaaA\kill

{\bf when} $p_i$ invokes ${\sf TO\_broadcast}(m)$ {\bf do}
$\ssend(m)$ $\tto$ itself.\\~\\

{\bf when} $m$ {\bf is received for the first time do} \\
\> $\bbroadcast(m)$;   $delivered_i\leftarrow delivered_i \cup\{m\}$.\\~\\

{\bf when} $(to\_deliverable_i$ contains messages not yet to-delivered)
{\bf do}\\
\>
{\bf let} $m$ = first message $\in to\_deliverable_i$ not yet to-delivered;\\
\>  ${\sf TO\_deliver}(m)$.\\~\\


{\bf background task} $T$ {\bf is}\\

\> {\bf repeat forever}\\

\> \>  $\wwait(delivered_i\setminus to\_deliverable_i \neq\emptyset)$;\\

\> \> {\bf let} $seq = (delivered_i\setminus to\_deliverable_i)$;\\
\> \>  order the messages in $seq$;\\

\> \> $sn_i\leftarrow sn_i +1$;
      $res_i \leftarrow CS[sn_i].{\sf propose}(seq)$;\\

\> \> add $res_i$ at the end of  $to\_deliverable_i$\\

\> {\bf end repeat}.

\end{tabbing}
\normalsize
\end{minipage}
}
\caption{TO-broadcast from consensus}
\label{fig:algo-tobroadcast-from-consensus}
}
\end{algorithm}

When a process $p_i$ invokes  ${\sf TO\_broadcast}(m)$
it sends the message to itself, which entails its broadcast, and only then
$p_i$  adds $m$ to  its  local set $delivered_i$. 
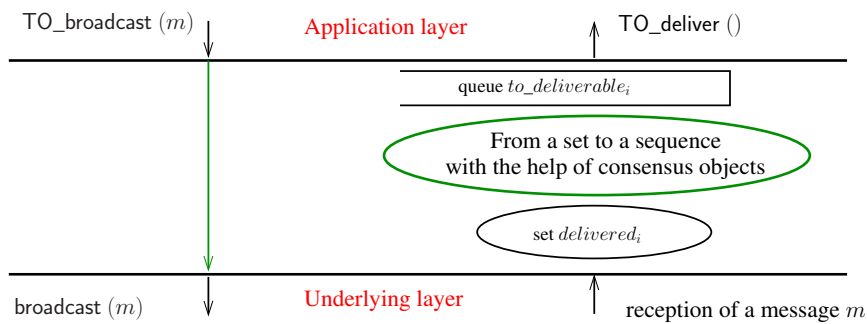
\begin{figure}[ht]
\centering{
\begin{minipage}[]{120mm}
\centering{
    \scalebox{0.35}{\input{fig-structure-from-cons-to-TO-URB.pstex_t}}
}
\end{minipage}
}
\caption{Structure of the consensus-based implementation of TO-broadcast}
\label{sidebar:tobroadcast-from-consensus}
\end{figure}
%
When a process receives a message $m$ from another process
for the first time, it does the same. It follows that when a process does
not crash during its broadcast of a message $m$, all processes receive it.  
Hence, if a process $p_j$ adds $m$ to $delivered_j$, so do at least
all the processes that do   not crash. 

When,  the queue $to\_deliverable_i$ of a process $p_i$
contains messages not yet locally to-delivered, $p_i$ to-delivers them
in the order in which they appear in  $to\_deliverable_i$.

The core of the algorithm is the background task $T$.
 A consensus object $SC[k]$ is associated with the
iteration number $k$. A process $p_i$ waits until there are messages
in the set $delivered_i$ and not yet in the queue $to\_deliverable_i$.
When this occurs, process $p_i$ computes this set of messages ($seq$)
and order them. Then it proposes $seq$ to the consensus instance
$SC[k]$.  This instance returns a sequence saved in $res_i$, which is
added by $p_i$ at the end of its local queue $to\_deliverable_i$.
The correctness of this algorithm relies on the properties of the
consensus object. For any $k\geq 1$, 
the consensus instance $CS[k]$ returns the same sequence
of messages to all the processes that invoke it.
As processes execute instances in the same order, their
queue $to\_deliverable_i$ eventually contain the same sequence of
messages. Formal proofs of this algorithms can be found in~\cite{CT96,R18}.

\begin{sidebar}[ht]
\fbox{
\centering{
\begin{minipage}[]{150mm}
\vspace{0.2cm}
While their styles are  different, these two citations capture
the universality issues
encountered in asynchrnonous fault-tolerant distributed  computing.  
\begin{itemize}
\item 
{\it In sequential systems, computability is understood through the
  Church-Turing Thesis: anything that can be computed,can be computed
  by a Turing Machine.  In distributed systems, where computations
  require coordination among multiple participants, computability
  questions have a different flavor. Here, too, there are many
  problems which are not computable, but these limits to computability
  reflect the difficulty of making decisions inthe face of ambiguity,
  and have little to do with the inherent computational power of
  individual participants.}

 Herlihy M., Rajsbaum S., and Raynal M.,
Power and limits of distributed computing shared memory models.
Theoretical Computer Science, 509:3-24 (2013).
\vspace{-0.2cm}
\item 
  {\it
    A distributed system is one in which the failure of a computer
    you didn't even know existed can render your own computer unusable.}

   L. Lamport,   email  Message-Id: <8705281923.AA09105@jumbo.dec.com>.\\

\end{itemize}
  
\end{minipage}
}
\caption{Two citations on universality}
\label{sidebar:citations}
}
\end{sidebar}

\section*{WHEN ARE UNIVERSAL CONSTRUCTIONS POSSIBLE?}

\paragraph{An impossibility.}
A fundamental result
in distributed computing is the impossibility to design a
(deterministic) algorithm that solves consensus in the presence of
asynchrony, even if only one process may crash, either in
message-passing~\cite{FLP85} or read/write shared memory systems~\cite{LA87}.
Given that consensus and TO-broadcast are equivalent, the state
machine replication algorithm presented above cannot be implemented in
asynchronous systems where processes can crash.

Thus, sequential thinking
for concurrent computing has studied properties about the underlying
system that enable the approach to go through.  There are several ways
of considering computationally stronger (read/write pr
message-passing) models (see, e.g.~\cite{R13,R18}), where state
machine replication can be implemented. Some ways, mainly suited to
message-passing systems, are presented in Sidebar~\ref{sidebar:omega}.
We discuss next a different way, through  powerful communication hardware.


\paragraph{The case of enriched read/write systems.}
Nearly all read/write systems usually provide processes with
synchronization-oriented atomic operations such as Test\&Set,
Compare\&Swap, or the pair of operations Load Link/Store Conditional
(LL/SC in short).  These operations have a consensus number greater
than $1$.  More specifically, the consensus number of Test\&Set is
$2$, while the consensus number of both Compare\&Swap and the pair
LL/SC, is $+\infty$.  Namely, $2$-process (but not a $3$-process)
consensus can be implemented from Test\&Set, despite crash failures.
Compare\&Swap (or LL/SC) can implement consensus for any number of
processes.  Hence, for any $n$, any object can be implemented in an
asynchronous $n$-process read/write system enriched with Compare\&Swap
(or LL/SC), despite up to $n-1$ process crashes.  Furthermore, that
are implementations that tolerate arbitrary, malicious
failures~\cite{C11,R18}.

\begin{sidebar}[h]
\fbox{
\centering{
  \begin{minipage}[t]{150mm}
   $~$\\ Ways of circumventing the consensus impossibility.
\begin{itemize}
  \vspace{-0.2cm}
\item  The failure detector approach can
\cite{CHT96} 
abstract away synchrony assumptions sufficient
to distinguish between slow processes and dead processes. 
\vspace{-0.2cm}
\item  In {\it eventually synchronous systems} 
  \cite{DDS87,DLS88} 
  there is a time after which the processes
  run synchronously. The celebrated Paxos algorithm is an example~\cite{L98}.
\vspace{-0.2cm}
\item By using {\it random} coins \cite{B83,R83}. 
  consensus is solvable with high probability.

\vspace{-0.2cm}
\item By using a synchronization operation with consensus number is $+\infty$ 
  if enriched read/systems.  
\end{itemize}
In the first three  cases, this means that the considered 
system is no longer fully asynchronous. 
\end{minipage}
}
\vspace{-0.5cm}
\caption{Circumventing consensus impossibility}
\label{sidebar:omega}
\vspace{-0.5cm}
}
\end{sidebar}

\paragraph{Consensus from the pair  LL/SC.}
The intuition of how the LL/SC operations work is as follows.
Consider a memory location $M$ accessed only by the operations LL/SC.
Assumed that if  a process invokes  $M.{\sf SC}(v)$ it has previously
invoked $M.{\sf LL}()$. 
The operation $M.{\sf LL}()$ is a simple read of $M$ which returns the
current value of $M$.  When  a process $p_i$ invokes $M.{\sf SC}(v)$
the value $v$ is written into $M$ if and only if no other process invoked
$M.{\sf SC}()$ since its ($p_i$) last invocation of  $M.{\sf LL}()$. 
If the write succeeds  $M.{\sf SC}()$ returns ${\tt true}$, otherwise it
returns ${\tt false}$ (see Sidebar~\ref{sidebar:LL-SC-operations}).

\begin{sidebar}[ht]
\fbox{
\centering{
\begin{minipage}[t]{150mm}
  {\centering{$~~~~~$ \scalebox{0.45}{\input{figure-LL-SC.pstex_t}}}} ~\\~\\
  Let $X$ and $Y$ be two different shared registers, and $p_i$, $p_j$, 
$p_k$ be three distinct processes. As there is no invocation of
$Y.{\sf SC}()$ between the invocations of $Y.{\sf LL}()$ and $Y.{\sf
  SC}()$ by $p_j$, its invocation of $Y.{\sf SC}()$ succeeds.
For the same reason, the invocation of for $X.{\sf SC}()$ by $p_i$ succeeds.
Differently, as there is an invocation of
$X.{\sf SC}()$ between the invocations of $X.{\sf LL}()$ and $X.{\sf
  SC}()$ by $p_k$, its invocation of $X.{\sf SC}()$ does not succeed.
\end{minipage}
}
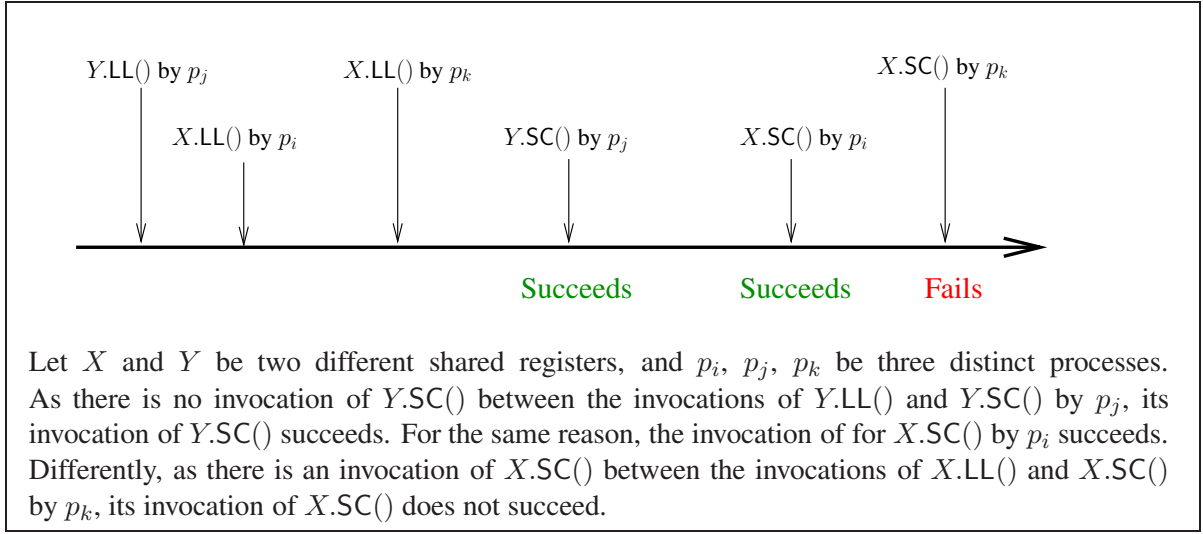
\caption{An execution of LL/SC operations}
\label{sidebar:LL-SC-operations}
}
\end{sidebar}

Algorithm~\ref{fig:algo-consensus-from-LL-SC} 
is a simple implementation of consensus  object
from the pair of operations LL/SC, which tolerates any number of
process crashes. 
The consensus object is represented by the memory location $M$
initialized to the default value $\bot$, which cannot be proposed).
Each process manages a local variable  $val_i$ and a Boolean $b_i$. 

\begin{algorithm}[h]
\centering{
\fbox{
\begin{minipage}[t]{150mm}
  \footnotesize
\renewcommand{\baselinestretch}{2.5}
\resetline
\begin{tabbing}
aaA\=aA\=aaaaaaaaaaaaA\kill

{\bf operation} ${\sf propose}(v)$ {\bf is}\\

\>   $val_i \leftarrow M.{\sf LL}()$;\\

\> {\bf if} $(val_i\neq \bot)$  \= {\bf then} \= ${\sf return}(val_i)$\\

\>   \> {\bf else} \> $b_i \leftarrow M.{\sf SC}(v)$;\\

\>   \>  \>  {\bf if} $b_i$ \= {\bf then} \=  ${\sf return}(v)$\\
\>   \>  \> \>  {\bf else} \>
    $val_i \leftarrow M.{\sf LL}()$; ${\sf return}(val_i)$ \\

\>   \>  \>  {\bf end if}\\

\> {\bf end if}.

\end{tabbing}
\normalsize
\end{minipage}
}
\vspace{-0.2cm}
\caption{Consensus from the operations LL/SC}
\label{fig:algo-consensus-from-LL-SC}
}
\end{algorithm}

When a process $p_i$ invokes the operation  ${\sf propose}(v)$ 
it first reads the value of $M$ (first invocation of  $M.{\sf LL}()$)
from which it obtains a value $val_i$.
If $val_i\neq \bot$, it is the value decided by the consensus object and
$p_i$ returns it. 
If $val_i= \bot$, no value has yet been decided and possibly several
processes are competing to impose their proposal as the decided value. 
Each of them invokes  $M.{\sf SC}()$.
Due the semantics of the pair LL/SC one and only one of them succeeds.
The winner returns its value, and the other competing processes
read again the value of $M$ (second invocation of  $M.{\sf LL}()$)
and return the value proposed by the winner.

\paragraph{A simple stacking-based universal construction.}
As consensus objects can be built from the pair of
operations LL/SC (Algorithm~\ref{fig:algo-consensus-from-LL-SC}) 
and  TO-broadcast communication abstraction can be built on top of
consensus objects   (Algorithm~\ref{fig:algo-tobroadcast-from-consensus}), 
their stacking allows us to use the universal construction
Algorithm~\ref{fig:TO-URB-based-universal-construction}
to obtain an implementation of any sequentially-defined object, 
which copes with the net effect of asynchrony and process failures.
This construction can give the reader  a feeling for
the distributed ledgers discussed in the next section.

\paragraph{A direct universal construction}
 Algorithm~\ref{fig:simple-LLSC-SM-universal-construction} (based on
 an algorithm introduced in~\cite{FK14}, simplified in~\cite{R17}) is
 a direct universal construction (does not use an intermediate layer
 of TO-broadcast) of an object $O$ with transition function $\delta$,
 for $n$ processes.

The shared memory is composed of the two  following data structures.
\begin{itemize}
\vspace{-0.2cm}
\item 
An array on $n$
atomic single-writer multi-reader registers, $\BOARD[1..n]$.
 While any process can read $\BOARD[i]$, only process $p_i$ can write it.
Each register $\BOARD[i]$ is composed of two fields:
$\BOARD[i].op$ which contains the last object operation invoked by $p_i$
and $\BOARD[i].sn$ which contains the associated local sequence number. 
\vspace{-0.2cm}
\item 
An atomic register,  $\STATE$,
accessed with the operations ${\sf LL}()$ and ${\sf SC}()$.
It is made of three fields: $\STATE.value$ contains
 the current state of the object under construction,
$\STATE.sn[1..n]$  is an array of local sequence number, and
$\STATE.res[1..n]$ which is an array of results.  More precisely,
$\STATE.res[i]$ contains the result of the last object operation
issued by $p_j$, and $\STATE.sn[j]$ contains its sequence number.
\end{itemize}
\begin{algorithm}[h]
\centering{
\fbox{
\begin{minipage}[t]{100mm}
\footnotesize
\renewcommand{\baselinestretch}{2.5}
\resetline
\begin{tabbing}
aaA\=aaA\=aaaaaaaaaaaaA\kill

{\bf when the operation} ${\sf op}_x ~(param_x)$ 
     {\bf is locally invoked}  {\bf do}\\

\> $sn_i \leftarrow sn_i+1$;
    $\BOARD[i] \leftarrow \langle {\sf op}_x ~(param_x),sn_i \rangle$;\\

\>  ${\sf apply}()$;\\

\>  $state_i \leftarrow \STATE.{\sf LL}()$;
     ${\sf return}(state_i.res[i])$.\\~\\

{\bf internal procedure} ${\sf apply}()$ {\bf is}\\

\> $state_i \leftarrow \STATE.{\sf LL}()$;\\



\> $board_i \leftarrow [\BOARD[1],\BOARD[2], \cdots,\BOARD[n]]$;\\

  , 
\> {\bf for} $\ell \in \{1, \cdots,n\}$ {\bf do}\\

\> \>  {\bf if}  \= $(board_i[\ell].sn = state_i.sn[\ell]+1)$\\

\> \> \> {\bf then} \=
      $\langle state_i.value, state_i.res[\ell]\rangle \leftarrow
                   \delta(state_i.value,pairs_i[\ell].op)$; \= \% line A\\

\>\>\>\> $~$

$state_i.sn[\ell]\leftarrow  state_i.sn[\ell]+1$  \> \% line B \\

\> \>  {\bf end if} \\

\> {\bf end for};\\

\> $success \leftarrow \STATE.{\sf SC}(state_i)$;\\

\>  {\bf if} $(\neg success)$
    {\bf then} \\

    \>   \>   $state_i \leftarrow \STATE.{\sf LL}()$;\\

\> \> {\bf if} $(sn_i =  state_i.sn[i]+1)$\\
\> \> \>  {\bf then} \=same as lines A and B with $\ell=i$;\\
\> \>\> \>                $\STATE.{\sf SC}(state_i)$ \\

\> \> 
{\bf end if}\\
   
\>  {\bf end if}.    
\end{tabbing}
\normalsize
\end{minipage}
}
\caption{Universal construction for  LL/SC-enriched shared memory systems
    (code for $p_i$)}
\label{fig:simple-LLSC-SM-universal-construction}
}
\end{algorithm}

Each process $p_i$ manages a local sequence number $sn_i$ and two
local variables, denoted $board_i$ and $state_i$, which will contain
local copies of $\BOARD$ and $\STATE$, respectively.

When a process $p_i$ invokes an operation  ${\sf op}_x ~(param_x)$ on $O$, 
it informs all the processes of it by storing the pair 
$\langle {\sf op}_x ~(param_x), sn_i\rangle$ in  $\BOARD[i]$.
It executes then  the internal procedure  ${\sf apply}()$
(which is the core of the construction). When it returns
from  ${\sf apply}()$, it returns the result that has been deposited in
$\STATE.res[i]$. As there is no waiting statement  in ${\sf apply}()$,
if the invoking process does not crash, it terminates its operation on $O$. 
Hence, the progress condition for object $O$ is wait-freedom.

When $p_i$ executes ${\sf apply}()$, if first atomically reads the
register $\STATE$ (invocation of  $\STATE.{\sf LL}()$), whose value
is saved in its local variable $state_i$,  reads the content of the
array $\BOARD$ and saves it in its local variable $board_i$.  Let us
remark that, while the the reading of each register $\BOARD[j]$ is
atomic, the array $\BOARD$ is read asynchronously and consequently
the reading of the whole array  $\BOARD$ is not at all atomic.
When this is done, $p_i$ starts a speculative execution, which
consists in a ''for'' loop,  with one iteration per process $p_\ell$. 
If the last operation announced by $p_\ell$ is the next to be applied
(according to its view of $p_\ell$'s local sequence numbers), $p_i$  applies
$p_\ell$'s operation to its local view of the current state of $O$,
namely $state_i$. When this has been done for each process $p_\ell$,
$p_i$ tries to write the new resulting state in  $\STATE$
(this is done by the invocation of $\STATE.{\sf SC}(state_i)$). 
If $\STATE.{\sf SC}(state_i)$ returns ${\tt true}$,
the speculative execution succeeded: $p_i$'s operation has been executed,
as have also  been  operations from other processes, and consequently
$p_i$'s invocation of ${\sf apply}()$ terminates.
Otherwise,  the speculative execution failed. 
In this case, process $p_i$ reads again $\STATE$
(second invocation of $\STATE.{\sf LL}()$). If its operation has not been
executed, $p_i$ speculatively  executes it on $state_i$, and
tries to commit it by invoking $\STATE.{\sf SC}(state_i)$. 
If this invocation returns ${\tt true}$ its operation is taken
into account. If it  returns ${\tt false}$, another process
$p_k$ invoked successfully  $\STATE.{\sf SC}(state_k)$
between the invocations of  $\STATE.{\sf LL}()$ and  $\STATE.{\sf SC}()$
by $p_i$. But in this case, due to the fact that LL/SC are atomic operations, 
necessarily when  $p_k$ read $\BOARD[i]$ it was informed of $p_i$'s
operation and consequently executed it. Hence, the result obtained
by $p_i$ from $\STATE.res[i]$ is the one associated with its last operation.


\section*{DISTRIBUTED LEDGERS}

Since ancient times, ledgers have been at the heart of
  commerce, to represent concurrent transactions by a permanent list
  of individual records sequentialized by date
  (Fig.~\ref{fig:ledger-object}).  Today we are
  beginning to see algorithms that enable the collaborative creation
  of digital distributed ledgers with properties and capabilities that
  go far beyond traditional physical ledgers.  All participants within
  a network can have their own copy of the ledger. Any of them can
  append a record to the ledger, which is then reflected in all copies
  in minutes or even seconds.  The records stored in the ledger can stay
  temper-proof, using cryptographic techniques.

\begin{figure}[h!]
\centering{
\begin{minipage}[]{150mm}
\centering{\scalebox{0.30}{\input{fig-blockchain.pstex_t}}}
\end{minipage}
}
\caption{Ledger object: a crucial  issue for the processes is to agree
  on the next block to add}
\label{fig:ledger-object}
\end{figure}
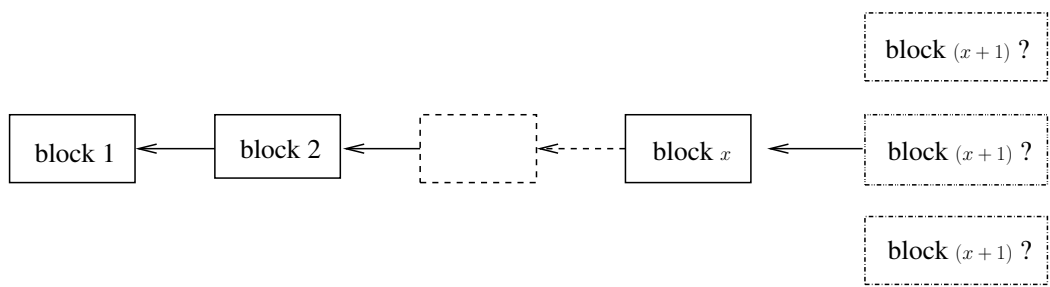

\paragraph{Ledgers as universal constructions.}
Mostly known because of their use in cryptocurrencies, and due to its
 \textit{blockchain} incarnation~\cite{N08}, from the
perspective of this paper a \textit{distributed ledger} is a byzantine
fault-tolerant replicated implementation of a specific {\it ledger}
object. The ledger object has two operations, ${\sf
  read}()$ and ${\sf append}()$.  Its sequential specification defines
it as a list of blocks.  A block $X$ can be added at the end of the
list with the operation ${\sf append}(X)$, while a ${\sf read}()$
returns the whole list. In the case of a cryptocurrency, $X$ may
contain a set of transactions.

Thus, a ledger object, as any other  object, can
be implemented in a distributed, fault-tolerant way, using the state
machine replication technique.  Furthermore, it can then be used as a
universal construction of an object $O$ defined by a state machine
with a transition function $\delta$. 
To do so, when a process invokes ${\sf append}(X)$, $X$ consists of a
transition to be applied to the state machine.
The state of the object is obtained through a ${\sf read}()$
invocation, which returns  the sequence of operations which have
been sequentially appended to the ledger, and then locally applying
them starting from the initial state of the object (see~\cite{R18} for
more details).
  
\paragraph{Three remarkable properties.}
The apparently innocent idea of a
${\sf read}()$ operation that returns the list of commands that have
been applied to the state machine,  opens the discussion of one of
the remarkable points of distributed ledgers that has brought them to
such wide attention. The possibility of guaranteeing a
\textit{temper-proof} list of commands. The blockchain implementation
is by using cryptographic hashes that link each record to the previous
one (although it actually has been known in cryptography community for
years~\cite{Merkle87}).

The ledger implementation used in Bitcoin showed that it is possible
to have a state machine replication tolerating Byzantine failures that
\textit{scales} to hundreds of thousands of processes.
The cost is temporality
sacrificing consistency--- forks can happen at the end of the
blockchain, which means that the last few records in the blockchain
may have to be withdrawn.

The third remarkable property brought to the public attention by
distributed ledgers is the issue of who the participants can be.  As
opposed to classic algorithms for mastering concurrency through
sequential thinking, the participants do not have to be a
priori-known, can vary with time, and may even be anonymous. Anyone
can append a block, and read the blockchain (although there are also
permissioned versions where participants have to be registered, and
even hybrid models).  In a sense, a distributed ledger is an open
distributed database, with no central authority, where the data itself
is distributed among the participants.

\paragraph{Agreement in dynamic systems.}
Bitcoin's distributed ledger implementation is relatively simple to
explain in the framework of state machine replication.  Conceptually
it builds on randomized consensus (something that had already been
carefully studied in traditional approaches,
e.g. Sidebar~\ref{sidebar:omega}), through the following ingenious
technique to implement it. Whenever several processes want to
concurrently append a block, they participate in a lottery. Each
process selects a random number (by solving cryptographic puzzles)
between $0$ and some large integer $K$, and the one that gets a number
smaller than $k<<K$, wins, and has the right to append its desired
block.  The implementation details of the lottery (by a procedure
called \textit{proof of work}) are not important for this paper; what
is important here, is that processes cannot cheat by biasing the
random number they get.  Thus, with high probability only one
wins. However, from time to time, more than one process wins and a
\textit{fork} happens, with more than one block being appended at the
end of the blockchain.  Again, for the purpose of this paper, it
suffices to say that only one branch eventually pervades (in Bitcoin
this is achieved by always appending to the longest branch).  This
introduces a new interesting idea into the paradigm of mastering
concurrency through sequential thinking: a tradeoff between faster
state machine replication, and temporary loss of consistency.
In other words, the $x$ operations at the very end of the blockchain,
for some constant $x$ (which depends on the 
assumptions about the environment) cannot yet be considered committed.
To be sure (with high probability) that an operation has permanently
been applied to the blockchain, a process has to wait until it is at a
depth greater than $x$ in the list of blocks.

\section*{ON THE LIMIT OF THE APPROACH}

It is intuitively clear, and it has been formally proved since a long
time that linearlizability is an expensive requirement.  Recent papers
in the context of shared memory programming, argue that it is often
possible to improve performance of concurrent data structures by
relaxing their semantics (see,
e.g.~\cite{CCBA17,HHHKLPSSV16,S11,TW18,TPKBAA13}).  In the context of
distributed systems, \textit{eventual consistency} is widely deployed
to achieve high availability by guaranteeing that if no new updates
are made to a given data item, eventually all accesses to that item
will return the last updated value~\cite{V09}.  Eventual consistency
(also called optimistic replication), which is deployed in some
distributed systems, has origins in early mobile computing. A system
that has achieved eventual consistency is often said to have
converged.  In the case of distributed ledgers, we have seen the
benefit that can be gained by relaxing the sequential approach to
mastering concurrency: branches at the end of the blockchain (such as
Bitcoin) temporarily violate a consistent view of the ledger.  Still,
blockchains suffer from a performance bottleneck due to the
requirement of ordering all transactions in a single list, which has
prompted the exploration of partially ordered ledgers, based on
directed acyclic graphs such as those based on Iota, Tangle, or Hedera
Hashgraph systems.  The benefit is scalability to thousands of
processes, that instead of communicating with each other to decide on
a single leader that will append a block, they avoid communication
altogether, using random numbers.

The \textit{CAP Theorem} formalizes a fundamental limitation of the
approach of mastering concurrency through sequential reasoning: at
most two of the following three properties are achievable, Consistency
(linearizability), Availability, Partition tolerance~\cite{GL02,GL12}.
This may give an intuition of why
distributed ledgers implementations have temporary forks.  An
alternative is a cost in availability, and postpone the property that
every non-failing participant returns a response for all operations
in a reasonable amount of time.  We have already seen in the ABD
algorithm that  
the system continues to function and upholds its
consistency guarantees, provided that only a minority of processes may fail.

Finally, another fundamental limitation to the approach of mastering
concurrency through sequential reasoning is that not all concurrent
problems of interest have sequential specifications.  Many examples
are discussed in~\cite{CRR18}, where a generalization of
linearizability to arbitrary concurrent specifications is proposed.


\section*{CONCLUSION}
The aim of this article was to show how does the theme of reducing
concurrent programming to sequential reasoning weaves through history
since the early days and along different domains (shared memory,
databases, distributed systems, cryptocurrencies, etc), to build
complex concurrent systems.  The thread brings in a scientific
foundation through common conceptual tools, such as sequential
specifications, progress and consistency conditions, synchronization
abstractions like consensus, communication mechanisms such as
broadcast and gossiping, fault-tolerance techniques, etc.  It evolves
from concrete resource-oriented mutual exclusion in a failure
free-context, through immaterial objects and failures, to universal
constructions of replicated state machines, to current trends on
dynamic, temper-proof distributed ledgers.  The deep continuity
lasting more than fifty years, is now exploring its frontiers, looking
for roundabouts to the inherent limitations of the approach.


\section*{ACKNOWLEDGMENTS}
This work has been  supported by the following  projects: French ANR 
16-CE40-0023-03 DESCARTES, 
devoted to modular structures in distributed computing, 
INRIA-UNAM {\it \'Equipe Associ\'ee} 
LiDiCo (at the Limits of Distributed Computing), and  UNAM PAPIIT  IN109917.


}
\end{document}

%% file: fig-history-dates-long-V2.pstex_t
\begin{picture}(0,0)%
\includegraphics{fig-history-dates-long-V2.pstex}%
\end{picture}%
\setlength{\unitlength}{4144sp}%
\begingroup\makeatletter\ifx\SetFigFont\undefined%
\gdef\SetFigFont#1#2#3#4#5{%
  \reset@font\fontsize{#1}{#2pt}%
  \fontfamily{#3}\fontseries{#4}\fontshape{#5}%
  \selectfont}%
\fi\endgroup%
\begin{picture}(14189,18455)(-5009,-25170)
\put(-2204,-15721){\makebox(0,0)[lb]{\smash{{\SetFigFont{25}{30.0}{\rmdefault}{\mddefault}{\updefault}{\color[rgb]{0,0,0}Impossibility of asynchronous determinisitic  consensus }%
}}}}
\put(-2249,-17071){\makebox(0,0)[lb]{\smash{{\SetFigFont{25}{30.0}{\rmdefault}{\mddefault}{\updefault}{\color[rgb]{0,0,0}Fast mutual exclusion  \cite{L87}}%
}}}}
\put(-4904,-18871){\makebox(0,0)[lb]{\smash{{\SetFigFont{20}{24.0}{\rmdefault}{\mddefault}{\updefault}{\color[rgb]{0,0,0}1993, 1997}%
}}}}
\put(-2249,-19816){\makebox(0,0)[lb]{\smash{{\SetFigFont{25}{30.0}{\rmdefault}{\mddefault}{\updefault}{\color[rgb]{0,0,0}Shared memory on top of asynchronous message-passing systems }%
}}}}
\put(-2249,-18016){\makebox(0,0)[lb]{\smash{{\SetFigFont{25}{30.0}{\rmdefault}{\mddefault}{\updefault}{\color[rgb]{0,0,0}Wait-free synchronization \cite{H91} (DA 2003)}%
}}}}
\put(-2294,-13921){\makebox(0,0)[lb]{\smash{{\SetFigFont{25}{30.0}{\rmdefault}{\mddefault}{\updefault}{\color[rgb]{0,0,0}Asynchronous  randomized  Byzantine consensus \cite{B83,R83} (DA 2015)}%
}}}}
\put(-2249,-18871){\makebox(0,0)[lb]{\smash{{\SetFigFont{25}{30.0}{\rmdefault}{\mddefault}{\updefault}{\color[rgb]{0,0,0}Transactional memory \cite{HM93,ST97} (DA 2012)}%
}}}}
\put(-4499,-12976){\makebox(0,0)[lb]{\smash{{\SetFigFont{20}{24.0}{\rmdefault}{\mddefault}{\updefault}{\color[rgb]{0,0,0}1981}%
}}}}
\put(-4499,-13921){\makebox(0,0)[lb]{\smash{{\SetFigFont{20}{24.0}{\rmdefault}{\mddefault}{\updefault}{\color[rgb]{0,0,0}1983}%
}}}}
\put(-4499,-14821){\makebox(0,0)[lb]{\smash{{\SetFigFont{20}{24.0}{\rmdefault}{\mddefault}{\updefault}{\color[rgb]{0,0,0}1985}%
}}}}
\put(-1799,-20446){\makebox(0,0)[lb]{\smash{{\SetFigFont{25}{30.0}{\rmdefault}{\mddefault}{\updefault}{\color[rgb]{0,0,0}despite a minority of process crashes \cite{ABD95} (DA 2011)}%
}}}}
\put(-2249,-13021){\makebox(0,0)[lb]{\smash{{\SetFigFont{25}{30.0}{\rmdefault}{\mddefault}{\updefault}{\color[rgb]{0,0,0}Simplicity in  mutex algorithms \cite{P81}}%
}}}}
\put(-1574,-16261){\makebox(0,0)[lb]{\smash{{\SetFigFont{25}{30.0}{\rmdefault}{\mddefault}{\updefault}{\color[rgb]{0,0,0}in the presence of process crashes \cite{FLP85} (DA 2001)}%
}}}}
\put(-4499,-15676){\makebox(0,0)[lb]{\smash{{\SetFigFont{20}{24.0}{\rmdefault}{\mddefault}{\updefault}{\color[rgb]{0,0,0}1985}%
}}}}
\put(-4499,-17071){\makebox(0,0)[lb]{\smash{{\SetFigFont{20}{24.0}{\rmdefault}{\mddefault}{\updefault}{\color[rgb]{0,0,0}1987}%
}}}}
\put(-4499,-17971){\makebox(0,0)[lb]{\smash{{\SetFigFont{20}{24.0}{\rmdefault}{\mddefault}{\updefault}{\color[rgb]{0,0,0}1991}%
}}}}
\put(-4499,-19771){\makebox(0,0)[lb]{\smash{{\SetFigFont{20}{24.0}{\rmdefault}{\mddefault}{\updefault}{\color[rgb]{0,0,0}1995}%
}}}}
\put(-4499,-21301){\makebox(0,0)[lb]{\smash{{\SetFigFont{20}{24.0}{\rmdefault}{\mddefault}{\updefault}{\color[rgb]{0,0,0}1996}%
}}}}
\put(-2249,-22786){\makebox(0,0)[lb]{\smash{{\SetFigFont{25}{30.0}{\rmdefault}{\mddefault}{\updefault}{\color[rgb]{0,0,0}Scalability, accountability  \cite{N08} }%
}}}}
\put(-4904,-12076){\makebox(0,0)[lb]{\smash{{\SetFigFont{20}{24.0}{\rmdefault}{\mddefault}{\updefault}{\color[rgb]{0,0,0}1980, 1982}%
}}}}
\put(-2204,-14821){\makebox(0,0)[lb]{\smash{{\SetFigFont{25}{30.0}{\rmdefault}{\mddefault}{\updefault}{\color[rgb]{0,0,0}Liveness (progress condition)   \cite{AS85}  (DA 2018)}%
}}}}
\put(-2249,-21346){\makebox(0,0)[lb]{\smash{{\SetFigFont{25}{30.0}{\rmdefault}{\mddefault}{\updefault}{\color[rgb]{0,0,0}Weakest information on failures to solve consensus in the }%
}}}}
\put(-1799,-22021){\makebox(0,0)[lb]{\smash{{\SetFigFont{25}{30.0}{\rmdefault}{\mddefault}{\updefault}{\color[rgb]{0,0,0}presence of asynchrony and process crashes \cite{CHT96,CT96} (DA 2010)}%
}}}}
\put(-2249,-11986){\makebox(0,0)[lb]{\smash{{\SetFigFont{25}{30.0}{\rmdefault}{\mddefault}{\updefault}{\color[rgb]{0,0,0}Byzantine failures in synchronous systems \cite{LSP82,PSL80} (DA 2005)}%
}}}}
\put(-3149,-6946){\makebox(0,0)[lb]{\smash{{\SetFigFont{20}{24.0}{\rmdefault}{\mddefault}{\updefault}{\color[rgb]{0,0,0}$~$}%
}}}}
\put(-2294,-9466){\makebox(0,0)[lb]{\smash{{\SetFigFont{25}{30.0}{\rmdefault}{\mddefault}{\updefault}{\color[rgb]{0,0,0}Mutual exclusion from non-atomic read/write registers \cite{L74}}%
}}}}
\put(-4544,-7486){\makebox(0,0)[lb]{\smash{{\SetFigFont{20}{24.0}{\rmdefault}{\mddefault}{\updefault}{\color[rgb]{0,0,0}1965}%
}}}}
\put(-4499,-8476){\makebox(0,0)[lb]{\smash{{\SetFigFont{20}{24.0}{\rmdefault}{\mddefault}{\updefault}{\color[rgb]{0,0,0}1971}%
}}}}
\put(-4544,-9376){\makebox(0,0)[lb]{\smash{{\SetFigFont{20}{24.0}{\rmdefault}{\mddefault}{\updefault}{\color[rgb]{0,0,0}1974}%
}}}}
\put(-4859,-10321){\makebox(0,0)[lb]{\smash{{\SetFigFont{20}{24.0}{\rmdefault}{\mddefault}{\updefault}{\color[rgb]{0,0,0}1977, 1983}%
}}}}
\put(-2249,-10321){\makebox(0,0)[lb]{\smash{{\SetFigFont{25}{30.0}{\rmdefault}{\mddefault}{\updefault}{\color[rgb]{0,0,0}Concurrent reading and writing  \cite{L77}, \cite{P83}}%
}}}}
\put(-4499,-11176){\makebox(0,0)[lb]{\smash{{\SetFigFont{20}{24.0}{\rmdefault}{\mddefault}{\updefault}{\color[rgb]{0,0,0}1978}%
}}}}
\put(-2249,-11221){\makebox(0,0)[lb]{\smash{{\SetFigFont{25}{30.0}{\rmdefault}{\mddefault}{\updefault}{\color[rgb]{0,0,0}Distributed state machine   \cite{L78} (DA 2000)}%
}}}}
\put(-2204,-8521){\makebox(0,0)[lb]{\smash{{\SetFigFont{25}{30.0}{\rmdefault}{\mddefault}{\updefault}{\color[rgb]{0,0,0}Semaphores \cite{D71}}%
}}}}
\put(-2249,-7621){\makebox(0,0)[lb]{\smash{{\SetFigFont{25}{30.0}{\rmdefault}{\mddefault}{\updefault}{\color[rgb]{0,0,0}Mutual exclusion from atomic read/write registers \cite{D65}}%
}}}}
\put(-4454,-22786){\makebox(0,0)[lb]{\smash{{\SetFigFont{20}{24.0}{\rmdefault}{\mddefault}{\updefault}{\color[rgb]{0,0,0}2008}%
}}}}
\put(-2249,-23686){\makebox(0,0)[lb]{\smash{{\SetFigFont{25}{30.0}{\rmdefault}{\mddefault}{\updefault}{\color[rgb]{0,0,0}Distributed recursion  \cite{GR10,LSP82} }%
}}}}
\put(-4949,-23731){\makebox(0,0)[lb]{\smash{{\SetFigFont{20}{24.0}{\rmdefault}{\mddefault}{\updefault}{\color[rgb]{0,0,0}1982, 2010}%
}}}}
\put(-4994,-24676){\makebox(0,0)[lb]{\smash{{\SetFigFont{20}{24.0}{\rmdefault}{\mddefault}{\updefault}{\color[rgb]{0,0,0} 2011, 2016}%
}}}}
\put(-2249,-24631){\makebox(0,0)[lb]{\smash{{\SetFigFont{25}{30.0}{\rmdefault}{\mddefault}{\updefault}{\color[rgb]{0,0,0}Distributed universality  \cite{GG11,H91,RST16} }%
}}}}
\end{picture}%

%% file: fig-registre-atomique-exemple.pstex_t
\begin{picture}(0,0)%
\includegraphics{fig-registre-atomique-exemple.pstex}%
\end{picture}%
\setlength{\unitlength}{4144sp}%
\begingroup\makeatletter\ifx\SetFigFont\undefined%
\gdef\SetFigFont#1#2#3#4#5{%
  \reset@font\fontsize{#1}{#2pt}%
  \fontfamily{#3}\fontseries{#4}\fontshape{#5}%
  \selectfont}%
\fi\endgroup%
\begin{picture}(12637,5159)(121,-7710)
\put(2026,-7486){\makebox(0,0)[lb]{\smash{{\SetFigFont{20}{24.0}{\familydefault}{\mddefault}{\updefault}{\color[rgb]{0,0,0}Here $R=1$}%
}}}}
\put(7696,-7486){\makebox(0,0)[lb]{\smash{{\SetFigFont{20}{24.0}{\familydefault}{\mddefault}{\updefault}{\color[rgb]{0,0,0}Here $R=2$}%
}}}}
\put(4501,-7486){\makebox(0,0)[lb]{\smash{{\SetFigFont{20}{24.0}{\familydefault}{\mddefault}{\updefault}{\color[rgb]{0,0,0}Here $R=3$}%
}}}}
\put(136,-3526){\makebox(0,0)[lb]{\smash{{\SetFigFont{25}{30.0}{\familydefault}{\mddefault}{\updefault}{\color[rgb]{0,0,0}$p_1$ }%
}}}}
\put(136,-4606){\makebox(0,0)[lb]{\smash{{\SetFigFont{25}{30.0}{\familydefault}{\mddefault}{\updefault}{\color[rgb]{0,0,0}$p_2$ }%
}}}}
\put(136,-5731){\makebox(0,0)[lb]{\smash{{\SetFigFont{25}{30.0}{\familydefault}{\mddefault}{\updefault}{\color[rgb]{0,0,0}$p_3$ }%
}}}}
\put(2206,-3166){\makebox(0,0)[lb]{\smash{{\SetFigFont{20}{24.0}{\familydefault}{\mddefault}{\updefault}{\color[rgb]{0,0,0}$R.{\sf read}() \rightarrow 1$ }%
}}}}
\put(6391,-3166){\makebox(0,0)[lb]{\smash{{\SetFigFont{20}{24.0}{\familydefault}{\mddefault}{\updefault}{\color[rgb]{0,0,0}$R.{\sf read}() \rightarrow 2$ }%
}}}}
\put(1081,-4291){\makebox(0,0)[lb]{\smash{{\SetFigFont{20}{24.0}{\familydefault}{\mddefault}{\updefault}{\color[rgb]{0,0,0}$R.{\sf write}(1)$ }%
}}}}
\put(4321,-4291){\makebox(0,0)[lb]{\smash{{\SetFigFont{20}{24.0}{\familydefault}{\mddefault}{\updefault}{\color[rgb]{0,0,0}$R.{\sf write}(2)$ }%
}}}}
\put(4996,-5461){\makebox(0,0)[lb]{\smash{{\SetFigFont{20}{24.0}{\familydefault}{\mddefault}{\updefault}{\color[rgb]{0,0,0}$R.{\sf write}(3)$ }%
}}}}
\put(7966,-5416){\makebox(0,0)[lb]{\smash{{\SetFigFont{20}{24.0}{\familydefault}{\mddefault}{\updefault}{\color[rgb]{0,0,0}$R.{\sf read}() \rightarrow 2$ }%
}}}}
\put(901,-2806){\makebox(0,0)[lb]{\smash{{\SetFigFont{20}{24.0}{\familydefault}{\mddefault}{\updefault}{\color[rgb]{0,0,0}$~$ }%
}}}}
\put(9451,-6586){\makebox(0,0)[lb]{\smash{{\SetFigFont{20}{24.0}{\familydefault}{\mddefault}{\updefault}{\color[rgb]{0,0,0}Omniscient observer's}%
}}}}
\put(10306,-7261){\makebox(0,0)[lb]{\smash{{\SetFigFont{20}{24.0}{\familydefault}{\mddefault}{\updefault}{\color[rgb]{0,0,0} time line}%
}}}}
\end{picture}%

%% file: fig-new-old-inversion.pstex_t
\begin{picture}(0,0)%
\includegraphics{fig-new-old-inversion.pstex}%
\end{picture}%
\setlength{\unitlength}{4144sp}%
\begingroup\makeatletter\ifx\SetFigFont\undefined%
\gdef\SetFigFont#1#2#3#4#5{%
  \reset@font\fontsize{#1}{#2pt}%
  \fontfamily{#3}\fontseries{#4}\fontshape{#5}%
  \selectfont}%
\fi\endgroup%
\begin{picture}(13323,5233)(2011,-1619)
\put(2476,-421){\makebox(0,0)[lb]{\smash{{\SetFigFont{20}{24.0}{\familydefault}{\mddefault}{\updefault}{\color[rgb]{0,0,0}$p_1$}%
}}}}
\put(2476,704){\makebox(0,0)[lb]{\smash{{\SetFigFont{20}{24.0}{\familydefault}{\mddefault}{\updefault}{\color[rgb]{0,0,0}$p_2$}%
}}}}
\put(2431,-1501){\makebox(0,0)[lb]{\smash{{\SetFigFont{20}{24.0}{\familydefault}{\mddefault}{\updefault}{\color[rgb]{0,0,0}$p_3$}%
}}}}
\put(9181,2864){\makebox(0,0)[lb]{\smash{{\SetFigFont{20}{24.0}{\familydefault}{\mddefault}{\updefault}{\color[rgb]{0,0,0}The  phase 1 majority  quorum obtained by $p_2$}%
}}}}
\put(2026,2864){\makebox(0,0)[lb]{\smash{{\SetFigFont{20}{24.0}{\familydefault}{\mddefault}{\updefault}{\color[rgb]{0,0,0}The  phase 1 majority  quorum obtained by $p_1$}%
}}}}
\put(2836,2369){\makebox(0,0)[lb]{\smash{{\SetFigFont{20}{24.0}{\familydefault}{\mddefault}{\updefault}{\color[rgb]{0,0,0}contains the pair $(v,tstamp(v))$}%
}}}}
\put(10486,974){\makebox(0,0)[lb]{\smash{{\SetFigFont{20}{24.0}{\familydefault}{\mddefault}{\updefault}{\color[rgb]{0,0,0}$\rread2()$ }%
}}}}
\put(5626,-151){\makebox(0,0)[lb]{\smash{{\SetFigFont{20}{24.0}{\familydefault}{\mddefault}{\updefault}{\color[rgb]{0,0,0}$\rread1()$ }%
}}}}
\put(9271,2369){\makebox(0,0)[lb]{\smash{{\SetFigFont{20}{24.0}{\familydefault}{\mddefault}{\updefault}{\color[rgb]{0,0,0}does not contain the pair $(v,tstamp(v))$}%
}}}}
\put(8326,-1321){\makebox(0,0)[lb]{\smash{{\SetFigFont{20}{24.0}{\familydefault}{\mddefault}{\updefault}{\color[rgb]{0,0,0}$\REG.\wwrite(v)$ }%
}}}}
\put(7831,3359){\makebox(0,0)[lb]{\smash{{\SetFigFont{20}{24.0}{\familydefault}{\mddefault}{\updefault}{\color[rgb]{0,0,0}$~$}%
}}}}
\end{picture}%

%% file: fig-structure-from-cons-to-TO-URB.pstex_t
\begin{picture}(0,0)%
\includegraphics{fig-structure-from-cons-to-TO-URB.pstex}%
\end{picture}%
\setlength{\unitlength}{4144sp}%
\begingroup\makeatletter\ifx\SetFigFont\undefined%
\gdef\SetFigFont#1#2#3#4#5{%
  \reset@font\fontsize{#1}{#2pt}%
  \fontfamily{#3}\fontseries{#4}\fontshape{#5}%
  \selectfont}%
\fi\endgroup%
\begin{picture}(14263,6435)(-3193,-3604)
\put(7156,1829){\makebox(0,0)[lb]{\smash{{\SetFigFont{25}{30.0}{\familydefault}{\mddefault}{\updefault}{\color[rgb]{0,0,0}${\sf TO\_deliver}~()$}%
}}}}
\put(1846,1784){\makebox(0,0)[lb]{\smash{{\SetFigFont{25}{30.0}{\familydefault}{\mddefault}{\updefault}{\color[rgb]{1,0,0}Application layer}%
}}}}
\put(1846,-2851){\makebox(0,0)[lb]{\smash{{\SetFigFont{25}{30.0}{\familydefault}{\mddefault}{\updefault}{\color[rgb]{1,0,0}Underlying  layer}%
}}}}
\put(-2924,1874){\makebox(0,0)[lb]{\smash{{\SetFigFont{25}{30.0}{\familydefault}{\mddefault}{\updefault}${\sf TO\_broadcast}~ (m)$ }}}}
\put(-854,2684){\makebox(0,0)[lb]{\smash{{\SetFigFont{12}{14.4}{\familydefault}{\mddefault}{\updefault}{\color[rgb]{0,0,0}$~$}%
}}}}
\put(7291,-3031){\makebox(0,0)[lb]{\smash{{\SetFigFont{25}{30.0}{\familydefault}{\mddefault}{\updefault}{\color[rgb]{0,0,0}reception of a message $m$}%
}}}}
\put(-3059,-2986){\makebox(0,0)[lb]{\smash{{\SetFigFont{25}{30.0}{\familydefault}{\mddefault}{\updefault}${\sf broadcast}~ (m)$ }}}}
\put(1801,-3481){\makebox(0,0)[lb]{\smash{{\SetFigFont{25}{30.0}{\familydefault}{\mddefault}{\updefault}$~$}}}}
\put(4231,-601){\makebox(0,0)[lb]{\smash{{\SetFigFont{25}{30.0}{\familydefault}{\mddefault}{\updefault}{\color[rgb]{0,0,0}with the help of consensus objects}%
}}}}
\put(4996,-151){\makebox(0,0)[lb]{\smash{{\SetFigFont{25}{30.0}{\familydefault}{\mddefault}{\updefault}{\color[rgb]{0,0,0}From a set to a sequence}%
}}}}
\put(4456,794){\makebox(0,0)[lb]{\smash{{\SetFigFont{20}{24.0}{\familydefault}{\mddefault}{\updefault}{\color[rgb]{0,0,0}queue $to\_deliverable_i$}%
}}}}
\put(5716,-1726){\makebox(0,0)[lb]{\smash{{\SetFigFont{20}{24.0}{\familydefault}{\mddefault}{\updefault}{\color[rgb]{0,0,0}set $delivered_i$}%
}}}}
\end{picture}%

%% file: figure-LL-SC.pstex_t
\begin{picture}(0,0)%
\includegraphics{figure-LL-SC.pstex}%
\end{picture}%
\setlength{\unitlength}{4144sp}%
\begingroup\makeatletter\ifx\SetFigFont\undefined%
\gdef\SetFigFont#1#2#3#4#5{%
  \reset@font\fontsize{#1}{#2pt}%
  \fontfamily{#3}\fontseries{#4}\fontshape{#5}%
  \selectfont}%
\fi\endgroup%
\begin{picture}(12868,3825)(1127,149)
\put(4726,3719){\makebox(0,0)[lb]{\smash{{\SetFigFont{20}{24.0}{\familydefault}{\mddefault}{\updefault}{\color[rgb]{0,0,0}$~$}%
}}}}
\put(9901,164){\makebox(0,0)[lb]{\smash{{\SetFigFont{25}{30.0}{\rmdefault}{\mddefault}{\updefault}{\color[rgb]{0,.56,0}Succeeds}%
}}}}
\put(7021,164){\makebox(0,0)[lb]{\smash{{\SetFigFont{25}{30.0}{\rmdefault}{\mddefault}{\updefault}{\color[rgb]{0,.56,0}Succeeds}%
}}}}
\put(2431,2189){\makebox(0,0)[lb]{\smash{{\SetFigFont{20}{24.0}{\familydefault}{\mddefault}{\updefault}{\color[rgb]{0,0,0}$X.{\sf LL()}$  by $p_i$}%
}}}}
\put(4681,3089){\makebox(0,0)[lb]{\smash{{\SetFigFont{20}{24.0}{\familydefault}{\mddefault}{\updefault}{\color[rgb]{0,0,0}$X.{\sf LL()}$  by $p_k$}%
}}}}
\put(6796,2189){\makebox(0,0)[lb]{\smash{{\SetFigFont{20}{24.0}{\familydefault}{\mddefault}{\updefault}{\color[rgb]{0,0,0}$Y.{\sf SC()}$  by $p_j$}%
}}}}
\put(9901,2189){\makebox(0,0)[lb]{\smash{{\SetFigFont{20}{24.0}{\familydefault}{\mddefault}{\updefault}{\color[rgb]{0,0,0}$X.{\sf SC()}$  by $p_i$}%
}}}}
\put(1306,3089){\makebox(0,0)[lb]{\smash{{\SetFigFont{20}{24.0}{\familydefault}{\mddefault}{\updefault}{\color[rgb]{0,0,0}$Y.{\sf LL()}$  by $p_j$}%
}}}}
\put(11701,3134){\makebox(0,0)[lb]{\smash{{\SetFigFont{20}{24.0}{\familydefault}{\mddefault}{\updefault}{\color[rgb]{0,0,0}$X.{\sf SC()}$  by $p_k$}%
}}}}
\put(12331,164){\makebox(0,0)[lb]{\smash{{\SetFigFont{25}{30.0}{\rmdefault}{\mddefault}{\updefault}{\color[rgb]{1,0,0}Fails}%
}}}}
\end{picture}%

%% file: fig-blockchain.pstex_t
\begin{picture}(0,0)%
\includegraphics{fig-blockchain.pstex}%
\end{picture}%
\setlength{\unitlength}{4144sp}%
\begingroup\makeatletter\ifx\SetFigFont\undefined%
\gdef\SetFigFont#1#2#3#4#5{%
  \reset@font\fontsize{#1}{#2pt}%
  \fontfamily{#3}\fontseries{#4}\fontshape{#5}%
  \selectfont}%
\fi\endgroup%
\begin{picture}(20541,6330)(-6332,-6169)
\put(5851,-286){\makebox(0,0)[lb]{\smash{{\SetFigFont{34}{40.8}{\rmdefault}{\mddefault}{\updefault}{\color[rgb]{0,0,0}$~$}%
}}}}
\put(6391,-3661){\makebox(0,0)[lb]{\smash{{\SetFigFont{34}{40.8}{\rmdefault}{\mddefault}{\updefault}{\color[rgb]{0,0,0}block $x$}%
}}}}
\put(10981,-1636){\makebox(0,0)[lb]{\smash{{\SetFigFont{34}{40.8}{\rmdefault}{\mddefault}{\updefault}{\color[rgb]{0,0,0}block $(x+1)$ ?}%
}}}}
\put(11026,-3661){\makebox(0,0)[lb]{\smash{{\SetFigFont{34}{40.8}{\rmdefault}{\mddefault}{\updefault}{\color[rgb]{0,0,0}block $(x+1)$ ?}%
}}}}
\put(11026,-5641){\makebox(0,0)[lb]{\smash{{\SetFigFont{34}{40.8}{\rmdefault}{\mddefault}{\updefault}{\color[rgb]{0,0,0}block $(x+1)$ ?}%
}}}}
\put(-5804,-3706){\makebox(0,0)[lb]{\smash{{\SetFigFont{34}{40.8}{\rmdefault}{\mddefault}{\updefault}{\color[rgb]{0,0,0}block 1}%
}}}}
\put(-1754,-3616){\makebox(0,0)[lb]{\smash{{\SetFigFont{34}{40.8}{\rmdefault}{\mddefault}{\updefault}{\color[rgb]{0,0,0}block 2}%
}}}}
\end{picture}%